\documentclass[aps,prb,twocolumn,floatfix]{revtex4-1}

\usepackage{graphicx}
\usepackage{amsfonts}
\usepackage{amsmath}
\usepackage{color}
\usepackage{bm}
\usepackage{amssymb}
\usepackage{wasysym}

\usepackage{multirow}
\usepackage{epstopdf}
\usepackage{natbib}
\usepackage{braket}

\usepackage{url}
\usepackage{soul}

\usepackage{xcolor}

\newcommand{\up}{\uparrow} 
\newcommand{\down}{\downarrow} 
\newcommand{\hp}[4]{ \begin{pmatrix}
  #1 & #2 \\
  #3 & #4 
\end{pmatrix}}

\def\eq#1{Eq.~(\ref{eq:#1})}
\def\fig#1{Fig.~(\ref{fig:#1})}

\begin{document}

\title{Efficient prediction of time- and angle-resolved photoemission Spectroscopy Measurements on a nonequilibrium BCS superconductor}
\author{Tianrui Xu, Takahiro Morimoto, Alessandra Lanzara and Joel E. Moore}
	\affiliation{Department of Physics, University of California, Berkeley, California 94720, USA}
	\affiliation{Materials Sciences Division, Lawrence Berkeley National Laboratory, Berkeley, California 94720, USA}

\pacs{}
\date{\today}

\begin{abstract}
We study how time- and angle-resolved photoemission (tr-ARPES) reveals the dynamics of BCS-type, s-wave superconducting systems with time-varying order parameters. Approximate methods are discussed, based on previous approaches to either optical conductivity or quantum dot transport, to enable computationally efficient prediction of photoemission spectra.  One use of such predictions is to enable extraction of the underlying order parameter dynamics from experimental data, which is topical given the rapidly growing use of tr-ARPES in studying unconventional superconductivity. The methods considered model the two-time lesser Green's functions with an approximated lesser self-energy that describes relaxation by coupling of the system to two types of baths. The approach primarily used here also takes into consideration the relaxation of the excited states into equilibrium by explicitly including the level-broadening of the retarded and advanced Green's functions. We present equilibrium and non-equilibrium calculations of tr-ARPES spectrum from our model and discuss the signatures of different types of superconducting dynamics.
\end{abstract}

\maketitle

\section{Introduction}
Angle-resolved photoemission spectroscopy (ARPES) is by now well established as a powerful technique to probe the electronic properties of a wide variety of solids~\cite{shenrmp03}. More recently, time-resolved ARPES (tr-ARPES) has been developed as a way to create and measure transient non-equilibrium states of a material that may not appear in its conventional phase diagram \cite{perfettiprl06,shensci08,perfettiaip12}.  In these experiments, an intense pulse ``pumps'' the system of interest into a non-equilibrium state, followed by a weak ``probe''. The ejected photo-electrons are then detected with energy and angle resolutions that depend on the time window of the probe pulse. Tr-ARPES measurements can achieve sufficient combined resolution of energy, momentum, and time evolution to study high-$T_c$ superconductors \cite{perfettiprl07,lanzaranat11,lanzarasci12,lanzara13prb,perfetti15prb,smallwood15prbr,dessau17,lanzaraprb18}, in addition to a variety of other materials\cite{wang2013observation,mahmood2016selective,smallwood16epl,lanzara16natcomm,gedik-vishik-review17}.

However, due to the non-equilibrium nature of these experiments, invariance under time translations is broken, with the consequence that one cannot simply analyze and interpret the experimental data through the usual, equilibrium formalism. While general methods that handle non-equilibrium systems, such as the Baym-Kadanoff-Keldysh non-equilibrium quantum field theory \cite{kamenev,rammerbook,gorshkov16prb}, exist in principle, applying these to predict tr-ARPES spectra that incorporate the transient dynamics of the system as well as the finite duration of the probe pulse is a nontrivial exercise~\cite{freericks09prl,freericks09prle,freericks15physcr}.

There have been a number of theoretical studies on dynamics of nonequilibrium superconductors \cite{rothwarftaylor67prl,volkov73jetp,schmid75,parker75prb,scalapino76prb,scalapino76prbe,howell04prl,axtkuhn04rpp,papenkort07prb,knorr08prb,freericks09prl,freericks09prle,zhu10prb,devereaux13prx,devereaux14prb,capone15prl,shen15prl,devereaux15prb,freericks15physcr,lanzara16prb,sentef16prb,devereaux17,demlersc17,murakami17,mazza17,mitra17prb,mitra18prl}. An example of the state of the art in numerical simulations of tr-ARPES experiments is Ref.~\onlinecite{devereaux17}, which shows how a sufficiently strong pump coupling to the electrons of a phonon-driven d-wave superconductor leads to amplitude (Higgs) mode oscillations at twice the mean gap frequency.  That work treats the superconducting gap self-consistently, i.e., changes in electron distribution induced by the pump modify the superconductivity, and as a result is computationally demanding even for a single pump strength/profile.  Here we will not treat the superconducting gap time evolution self-consistently; the goal is that by finding efficient means to compute how different gap evolutions and probe properties would lead to different tr-ARPES signals, our approach can be used to interpret tr-ARPES experimental data and learn how the underlying superconducting gap evolved.

In this respect, our work is more similar to recent theoretical models of a different problem, namely the time-dependent optical conductivity in pump-probe experiments on superconductors.  Complementary to work on tr-ARPES, sophisticated non-equilibrium methods have recently been applied to study the transient optical conductivity of non-equilibrium superconducting systems~\cite{millis17prb,millis17nat}. These works were motivated in part by the experimental observation that, in several kinds of superconductors~\cite{cavalleri16}, a strong pulse significantly modifies the reflectivity or transmissivity signal that at equilibrium is a standard probe of superconductivity.  As the interpretation of these signals in a non-equilibrium context can be subtle~\cite{orensteindodge}, a practical non-equilibrium theoretical approach must be developed to extract information from experimental data about the underlying nonequilibrium processes and states, which is a kind of ``inverse problem.'' For  example,  given  data,  what  is  the most likely time dependence of superconductivity to explain the observed results?  How does the effective damping depend on system parameters?

That task has been underway for some time in optical conductivity and is here undertaken for tr-ARPES.  The main aspect in which our formalism differs from the previous work on optical conductivity in superconductors, aside from being about a different measurement, is in the detailed treatment of dissipation or level broadening in the system. The importance of dissipation or level broadening can be seen from, for example, considering an idealized probe that suddenly changes (``quenches'') the electronic Hamiltonian from a metallic to a superconducting form. In the absence of dissipation, the fact that the metallic ground state is not the ground state of the new Hamiltonian means that some excited states will be occupied. These states would appear as positive-energy states in an tr-ARPES measurement that never dissipate, which does not seem to be what typically happens in reality, where the tr-ARPES intensity above the Fermi level tends to decay as time goes \cite{lanzaranat11,lanzarasci12,lanzara13prb,perfetti15prb,smallwood15prbr,dessau17}. 

Just as in the theories of pump-probe optical conductivity cited above, adding some form of dissipation is needed if the system is to return to equilibrium eventually.  Coupling the system to a ``bath'' of additional electronic states as in studies of optical conductivity will lead to relaxational time-dependence, but if the bath is finite (as needed for computational purposes) then eventually there will be oscillatory behavior rather than a return of the system to equilibrium.  Studying longer times requires larger baths to avoid such oscillations.  To prevent non-dissipating positive-energy states and/or oscillations, which could exist in tr-ARPES experiments in principle but do not seem to be observed, we adapt a method previously used to incorporate relaxation in the theory of non-equilibrium phenomena in quantum dots~\cite{Jauho}. This approach could be viewed as an approximation to a thermodynamically large bath whose treatment is computationally not feasible.  We comment in closing on some other possible uses and advantages of this approach.

The specific case treated here is the calculation of the tr-ARPES signal of a BCS s-wave superconducting system with a specified non-equilibrium superconducting order parameter $\Delta(t)$.  We use the mean-field (Bogoliubov-de Gennes) approximation to the superconducting system, together with the Keldysh formalism, to tackle the problem. As the approach is also feasible for general momentum and time dependence of the order parameter, it could be easily generalized to superconducting systems other than s-wave.  The formalism is computationally efficient enough to be used for the inverse problem, i.e., given an experimental tr-ARPES profile for one or more probe windows, one could compare it to different possible profiles or momentum dependences of the superconducting order parameter.

The rest of the paper is organized as follows. In Sec.~\ref{keldysh}, we review the theory of the tr-ARPES signal, as well as the Keldysh formalism that applies to such calculations.  In Sec.~\ref{glesser}, we present two approximations to the two-time lesser Green's function, which is the key building block of the tr-ARPES signal, and discuss the assumptions of each approximation. In Sec.~\ref{trcalc}, we show several tr-ARPES calculations with different temporal profiles of BCS order parameters. Finally we summarize and discuss future directions in Sec.~\ref{discuss}. 

\section{TR-ARPES Signal From Keldysh Formalism} \label{keldysh}

In this section, we review the Keldysh formalism applied to the Bogoliubov-de Gennes (BdG) description of a superconductor and the theory of tr-ARPES signals.  Then we explain how the lesser Green's function, which is crucial in simulating tr-ARPES signals, is calculated.

\subsection{Keldysh formalism in Bogoliubov-de Gennes (BdG) equation}

We first introduce Nambu-spinor notation:\cite{dgscbook,schriefferscbook,fu08}
\begin{equation}
  \Psi_k=(c_{k,\up},c_{k,\down},c^\dagger_{-k,\down}, -c^\dagger_{-k,\up})^T,
  \label{eq:nambu spinor}
\end{equation}
where $c^\dagger_{k,\sigma}$ creates an electron of momentum $k$ and spin $\sigma$. The superscript $T$ denotes transposition.
One can then write Green's functions in \eq{nambu spinor} basis as follows \cite{rammerbook,rammersmithrmp,aoki14rmp}:
\begin{align}
  G^R_{\alpha\beta}(t,t')&=-i\theta(t-t')\braket{\{\Psi_\alpha(t),\Psi_\beta^{\dagger}(t')\}}\nonumber\\
  G^A_{\alpha\beta}(t,t')&=i\theta(t'-t)\braket{\{\Psi_\alpha(t),\Psi_\beta^{\dagger}(t')\}}\nonumber\\
  G^K_{\alpha\beta}(t,t')&=-i\braket{[\Psi_\alpha(t),\Psi_\beta^{\dagger}(t')]}\nonumber\\
  G^<_{\alpha\beta}(t,t')&=i\braket{\Psi^{\dagger}_\alpha(t')\Psi_\beta(t)},
  \label{eq:grakldef}
\end{align}
where $R$, $A$, $K$, and $<$ stand for retarded, advanced, Keldysh, and lesser, and $\braket{\cdots}$ is taken with respect to the ground state of the system at zero temperature. 
The Green's functions are matrices acting on the Nambu spinor basis that are labeled by the indices $\alpha,\beta$. 


It is straightforward to obtain $G^{R/A}$ when the system is described by a Hamiltonian, $\mathcal{H}$, made of fermion bilinears, as is the case we focus on in this paper.
When the Hamiltonian $\mathcal{H}$ is given by
\begin{equation}
  \mathcal{H}(t)=\Psi^\dagger H(t)\Psi,
\end{equation}
where $H(t)$ is the Hamiltonian matrix,
$G^{R/A}$ are computed by solving the differential equation \cite{millis17prb,rammersmithrmp},
\begin{align}
  G^R(t,t)&=-i\nonumber\\
  i\partial_t G^R(t,t')&=H(t)G^R(t,t') \qquad t>t'\nonumber\\
  G^A(t,t')&=[G^R(t',t)]^\dagger.
  \label{eq:grga0}
\end{align}
Obtaining $G^{K/<}$ requires solving the Keldysh equation \cite{kamenev} as we explain in the following.

\subsection{TR-ARPES signal from lesser Green's function}
Once the lesser Green's function $G^<$ is obtained, the tr-ARPES signal can be calculated from \cite{freericks09prl}
\begin{align}
  I(k,\omega,t)\propto{\text {Im}}\int_{t_0}^tdt_1\int_{t_0}^t&dt_2G^<_k(t_1,t_2)\nonumber\\
    &\times s(t_1)s(t_2)e^{i\omega(t_1-t_2)}
  \label{eq:Ikwt}
\end{align}
where $s(t)$ is the temporal profile of the probe pulse, the integration limits, $t_0$ and $t$, are controlled by $s(t)$ as the probe pulse has finite width. And we have 
\begin{equation}
    G^<_{k,{\sigma\sigma'}}(t_1,t_2)=i\braket{c^{\dagger}_{k,\sigma}(t_2)c_{k,\sigma'}(t_1)},
    \label{eq:Glesser normal}
\end{equation}
which is the normal component of the $G^<$ defined in \eq{grakldef} with spin indices $\sigma, \sigma'$.

In our calculations for tr-ARPES signals that we show later, we consider systems with spin SU(2) symmetry. In this case, the BdG Hamiltonian \cite{dgscbook} is decoupled to two identical 2$\times$2 Hamiltonians spanned by $(c_{k,\up}, c_{-k\down}^\dagger)$ and $(c_{k,\down},- c_{-k\up}^\dagger)$, respectively. 
By focusing on the first one, we can suppress   the spin indices in \eq{Glesser normal}.

\subsection{Lesser Green's function} \label{glesser}
Now we explain how the lesser Green's function is computed in our framework.
We calculate $G^<$ by using Keldysh equation
in time domain \cite{pdan84}
\begin{equation}
  G^<(t,t')=\int dt_1\int dt_2G^R(t,t_1)\Sigma^<(t_1,t_2)G^A(t_2,t'),
  \label{eq:glkeldysh}
\end{equation}
where all indices but time are suppressed.
Here $\Sigma^<$ is the lesser self-energy that effectively determines occupation of electrons in energy eigenstates. Since $G^{R/A}$ are calculated from \eq{grga0}, one needs to specify the form of the lesser self-energy $\Sigma^<$. 

In our framework, we consider the system is coupled to the heat bath with large bandwidth. In this case, we can explicitly write  $\Sigma^<$ as \cite{Jauho}
\begin{align}
    \Sigma^<(t_1,t_2)&=i\gamma\int\frac{d\omega}{2\pi}f(\omega)e^{-i\omega(t_1-t_2)}\nonumber\\
    &=\frac{-\gamma/2\pi}{t_1-t_2+i0^+},
  \label{eq:slapprox}
\end{align}
by integrating out the heat bath,
where $\gamma$ is the level broadening of $G^{R/A}$.
This form of self energy introduces a  dissipation effect to the system (with the timescale of $\sim 1/\gamma$) as well as the level broadening of the energy eigenstates ($\sim \gamma$).

Thus we calculate lesser Green's functions via
\begin{align}
  &G^<_k(t,t')\nonumber\\
  &=\frac{-\gamma}{2\pi}\int dt_1\int dt_2\frac{G^R_k(t,t_1)G^A_k(t_2,t')}{t_1-t_2+i0^+}\nonumber\\
  &\qquad\qquad\qquad\times e^{-\gamma(t-t_1+t'-t_2)/2},
    \label{eq:glcal}
\end{align}
where $e^{-\gamma(\cdots)/2}$ is the explicit form of level broadenings of $G^{R/A}$.
In the equilibrium systems, \eq{slapprox} combined with \eq{glkeldysh} reproduces the correct form of $G^<(\omega)$ (see Appendix \ref{app:eqgl} for detailed derivations) after Fourier transformation. 

We note that Ref.~\onlinecite{millis17prb} used another approximation to calculate $G^<$:
\begin{equation}
  G^<_k(t_1,t_2)=iG^R_k(t_1,t_0)n_{0,k}G^A_k(t_0,t_2),
  \label{eq:glnodis}
\end{equation}
where $n_{0,k}=\braket{\Psi^{\dagger}_k\Psi_k}$ at the initial time $t_0$.
In the case of metals, this method gives the correct forms of lesser Green's function, and hence, tr-ARPES spectrum in the equilibrium.
It also gives the correct answer for equilibrium superconductors, if we use $n_{0,k}$ that is diagonal in the energy eigenstates of BdG Hamiltonian.
However, one major difference between these two formalisms is that a dissipation effect is directly incorporated in \eq{slapprox}, while \eq{glnodis} gives pure unitary dynamics of time-dependent system (dissipation results from the perspective of one part of the system, which is indeed the microscopic origin of real dissipation).

For example, if we consider a quench problem where a system evolves with 
\begin{equation}
  H(t)=\theta(-t)H_{metal}+\theta(t)H_{BCS},
  \label{eq:qhami}
\end{equation}
there emerges a quenched peak at positive energy, which is shown in Fig.\ref{fig:quenchpk} (see Appendix \ref{app:nondis} for details). With \eq{glnodis}, the weight in the positive tr-ARPES peak does not decay due to the absence of dissipation.
This coincides with the case of $\gamma \to 0$ in our formalism (see Appendix \ref{app:nondis} for detailed derivations).
To incorporate dissipation effects with \eq{glnodis}, one needs to explicitly couple the system to an external bath that dissipates the extra energy \cite{millis17prb}.  This is physically valid and has the advantage of allowing different microscopic dissipation mechanisms.  However, it is computationally advantageous to incorporate dissipation in $\Sigma^<$, so that one does not need to solve time evolution of both the system and the heat bath.

The approach using Eqs.(\ref{eq:glkeldysh}) and (\ref{eq:slapprox}) is used in other areas of non-equilibrium physics and facilitates numerical calculations, especially if the number of bath degrees of freedom is large.  As previously mentioned, simulating a heat bath explicitly may lead to unphysical oscillation of expectation values that depends on the system size of the heat bath.  

Finally, \eq{glnodis} is not applicable to interacting systems that involve, for example, four-fermion terms. This can be understood from the absence of dissipation since quasiparticles in interacting systems generally have a finite lifetime.  In the formalism using Eqs.(\ref{eq:glkeldysh}) and (\ref{eq:slapprox}), however, extension to interacting systems is straightforward if one uses $G^{R/A}$ and $\Sigma^<$ for the interacting systems in \eq{glkeldysh}, and incorporate the effect of heat bath by introducing a level broadening to $G^{R/A}$ and the corresponding self energy of $G^<$.

\begin{figure}
  \begin{center}
    \includegraphics[scale=0.4]{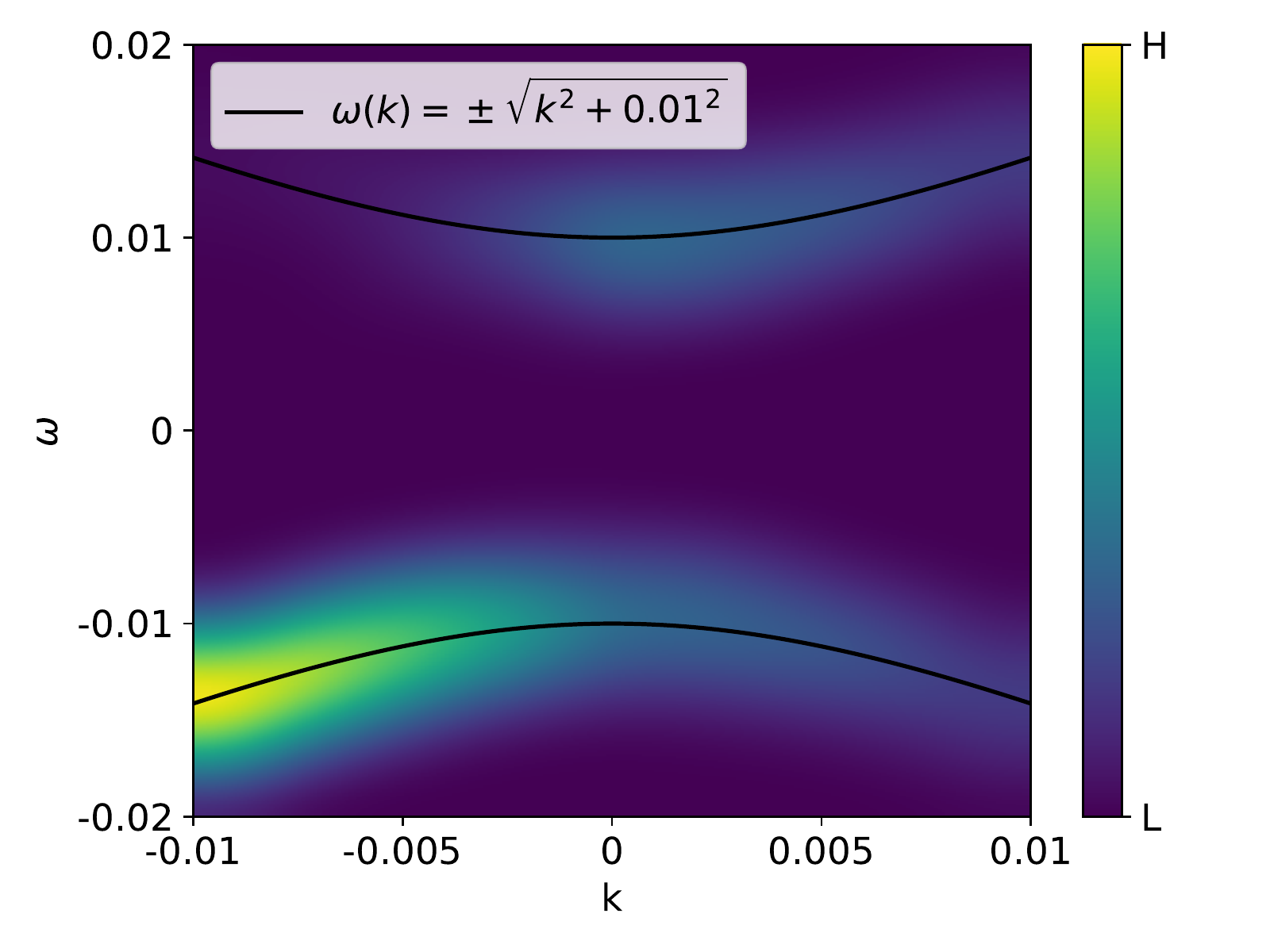}
    \caption{Non-equilibrium tr-ARPES calculation of constant superconducting gap with a quench, described by \eq{qhami} with a BCS superconducting gap $\Delta=0.01$. The lesser Green's function is calculated using \eq{glnodis} with $\epsilon_k=k$, $t_p=500$, $t_0=0$, $t=1000$, $\sigma=400$. The black curve shows $\omega(k)=\pm\sqrt{k^2+\Delta^2}$.}
    \label{fig:quenchpk}
  \end{center}
\end{figure}

\section{TR-ARPES Signal: Calculation} \label{trcalc}

In this section, we show calculations of tr-ARPES signals of systems with different temporal profiles of superconducting order parameter. We start the section by introducing the system that we are considering, both the hamiltonian of the system as well as the profile of our probe pulse. Then, we provide calculations of: metal, equilibrium BCS, and non-equilibrium BCS systems.

\subsection{Model of 1D superconductor}

We consider a BCS-type, s-wave superconducting system with time-varing order parameter:
\begin{equation}
  H=\sum_{k,\sigma}\epsilon_{k}c^{\dagger}_{k,\sigma}c_{k,\sigma}+\Delta(t)\sum_{k}c^{\dagger}_{k,\up}c_{-k,\down}+h.c.,
  \label{eq:hami}
\end{equation}
where $\Delta(t)$ is the {\it post-pump} time-dependent superconducting gap, which is an input to the theory, and is set to be real-valued for now. 

We do not consider how $\Delta(t)$ varies due to the time-dependence of the pump fields. For $\epsilon_k$, we consider a linear dispersion near fermi surface, i.e.: $\epsilon_k=k$.

Note that if the pump field is still on during the probe measurement, one needs to write the lesser Green's function into its gauge-invariant form \cite{freericks15physcr}:
\begin{equation}
    G_k^<(t_1,t_2)\to G_{\tilde{k}}^<(t_1,t_2),
\end{equation}
by shifting the momentum $k$ via
\begin{align}
    \tilde{k}&\to k+\frac{1}{t_1-t_2}\int_{-(t_1-t_2)/2}^{(t_1-t_2)/2}dt' A_{pump}\left(\frac{t_1+t_2}{2}+t'\right).
\end{align}

We rewrite the hamiltonian (\ref{eq:hami}) into Nambu-spinor basis:
\begin{equation}
  \mathcal{H}=\sum_{k}\Psi^{\dagger}_kH_k(t)\Psi_k,
  \label{eq:hamiN}
\end{equation}
where $H_k(t)$ takes the form of
\begin{equation}
  H_k(t)=\hp{\epsilon_{k}}{\Delta(t)}{\Delta(t)}{-\epsilon_{k}}.
  \label{eq:hk}
\end{equation}
Here we consider $2\times2$ BdG Hamiltonian spanned by the two-component spinor $\Psi_k=(c_{k,\up}, c^\dagger_{-k,\down})^T$, by assuming spin SU(2) symmetry.

We thus can solve $G^{R/A}_k$ from
\begin{align}
  G^R_k(t,t)&=-i\nonumber\\
  i\partial_tG^R_k(t,t')&=H_k(t)G^R_k(t,t') \qquad t>t'\nonumber\\
  G^A_k(t,t')&=[G^R_k(t',t)]^\dagger.
  \label{eq:grga}
\end{align}
We note two things regarding \eq{grga}. First, the above equations are applicable to the systems that are decoupled in $k$-space, where the retarded and advanced Green's function can be solved in each momentum separately. Second, the gap function can be regarded as an anomalous self-energy. In the BCS Hamiltonian that we employ, one assumes an energy independent order parameter (i.e., a $\delta$ - function in the time domain), which gives \eq{grga}. In strong coupling regime, on the other hand, the retardation of electron-phonon interaction can be modeled by two-time self-energy, where one needs to solve $G^R$ with the full time-dependence of its self-energy.

At the same time, we consider a simple, gaussian probe pulse:
\begin{equation}
  s(t)=\frac{1}{\sqrt{2\pi\sigma^2}}e^{-((t-tp)/\sigma)^2},
  \label{eq:probe}
\end{equation}
where $t_p$ is where the probe pulse centers, and $\sigma$ tunes the width of the probe pulse.

\subsection{Equilibrium superconducting system}
In this subsection, we demonstrate numerically that tr-ARPES signals in the equilibrium systems are reproduced by our formalism. 

We first present calculations of tr-ARPES signal of metals, i.e. $\Delta=0$ in \eq{hami}, from \eq{Ikwt}, with both $G^{R/A}$ and $G^<$ calculated numerically via Eqs.~(\ref{eq:grga}) and (\ref{eq:glcal})

\begin{figure}
  \begin{center}
    \includegraphics[scale=0.25]{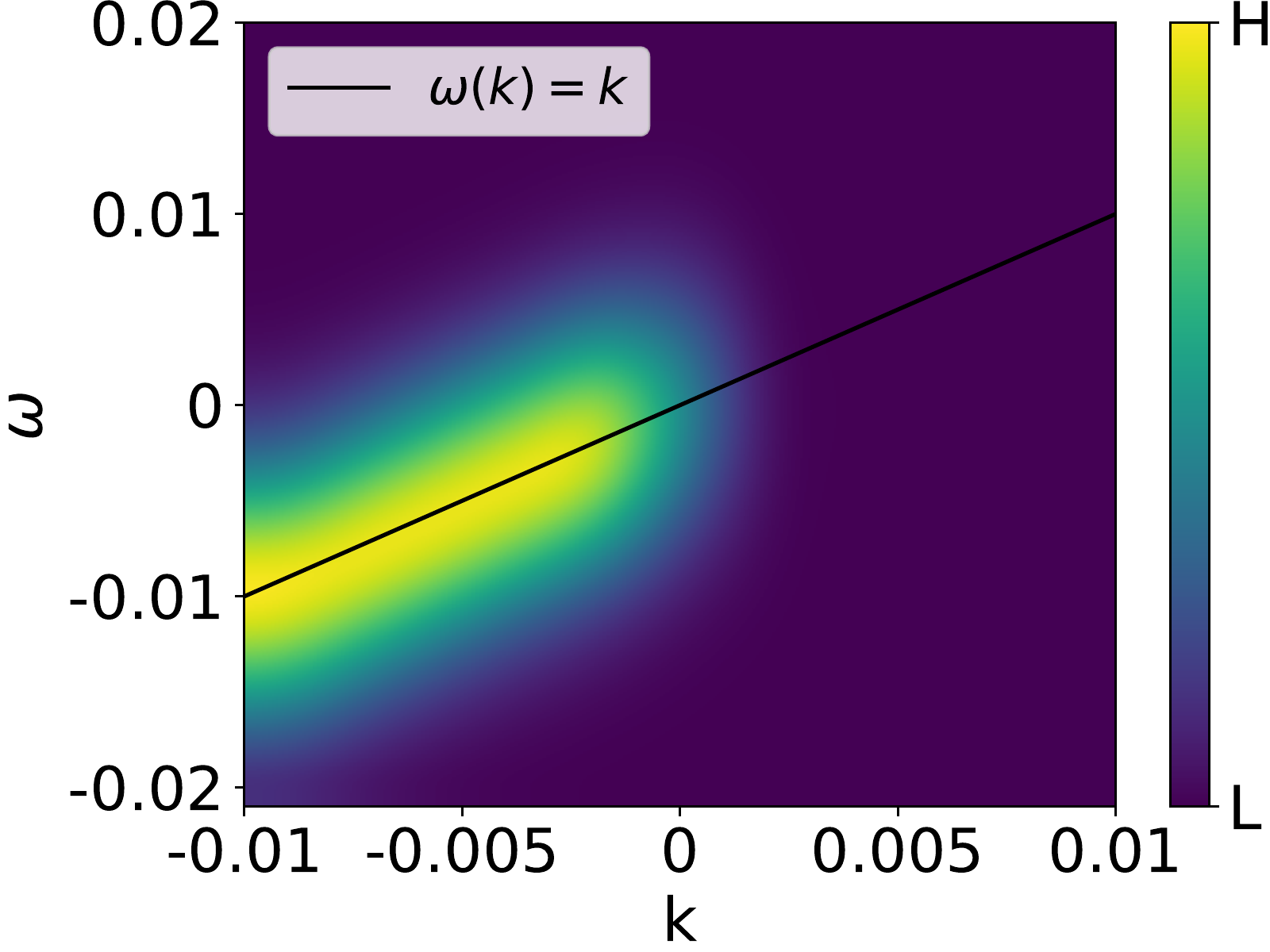}
    \includegraphics[scale=0.25]{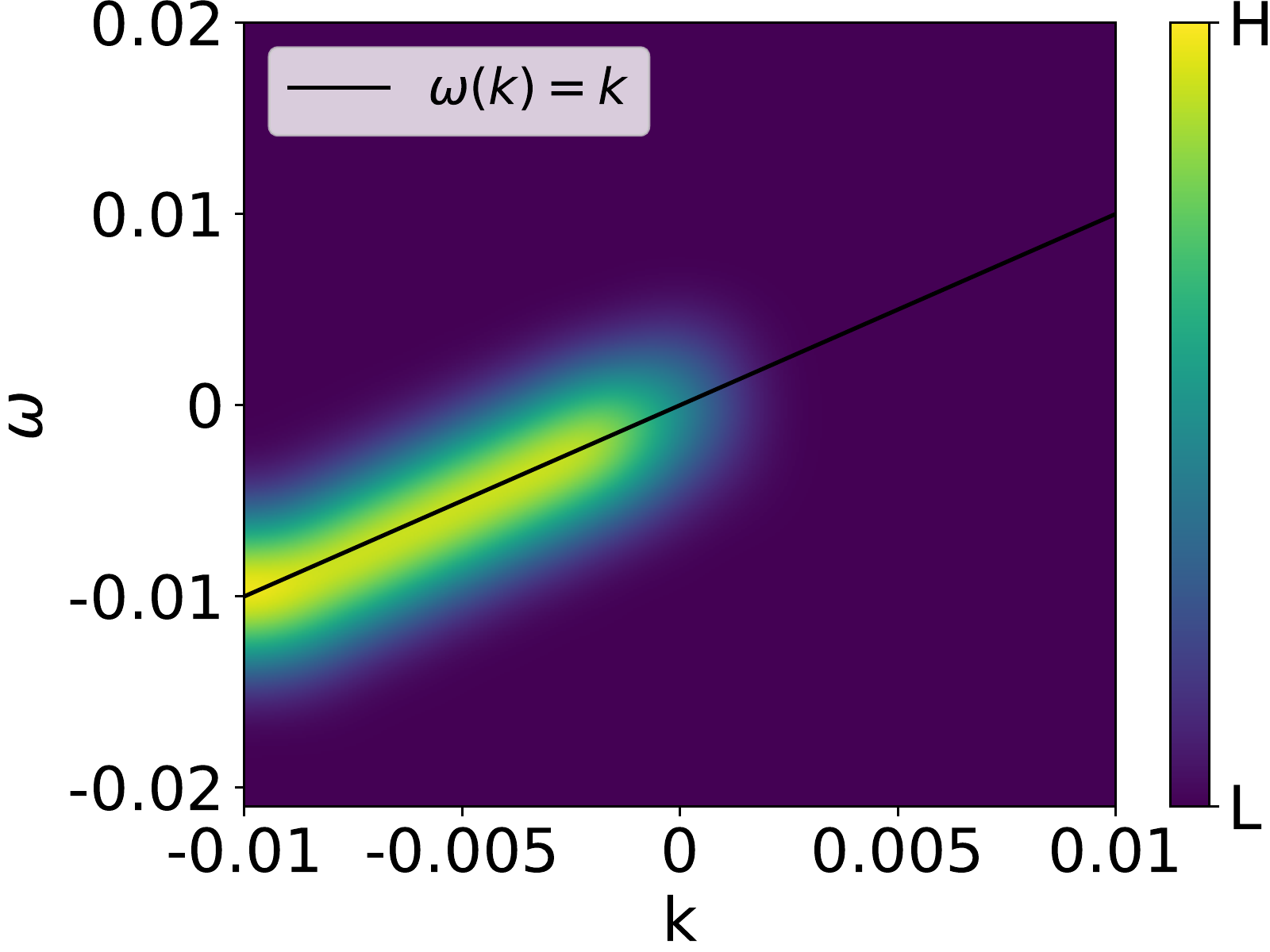}
    \caption{Equilibrium tr-ARPES calculations of metal, with $\epsilon_k=k$, $\gamma=0.0001$, $t_p=500$, $t=1000$, $\sigma=200$ (left panel) $\sigma=400$ (right panel). The black curve shows $\omega(k)=\epsilon_k$.}
    \label{fig:metal}
  \end{center}
\end{figure}

\fig{metal} shows two calculations of tr-ARPES signal of a metal, with numerical solutions to both $G^{R/A}$ from \eq{grga} and $G^<$ from \eq{glcal}. The only difference in between of the two panels are the width of the probe pulse, $\sigma$, which determines the energy resolution of the signals, the same as what was mentioned in Ref.~\onlinecite{devereaux13prx}. The trend that as $\sigma$ increases, the width of the tr-ARPES signal decreases, also indicates that as $\sigma\to\infty$, one may expect \cite{freericks09prl}
\begin{equation}
    I(k,\omega,t)\propto A(k,\omega)f(\omega),
\end{equation}
where $f(\omega)$ is the Fermi-Dirac distribution at the modeling temperature, in our case, $0K$.

Next, we show tr-ARPES signal of a superconductor.

\begin{figure}
  \begin{center}
    \includegraphics[scale=0.25]{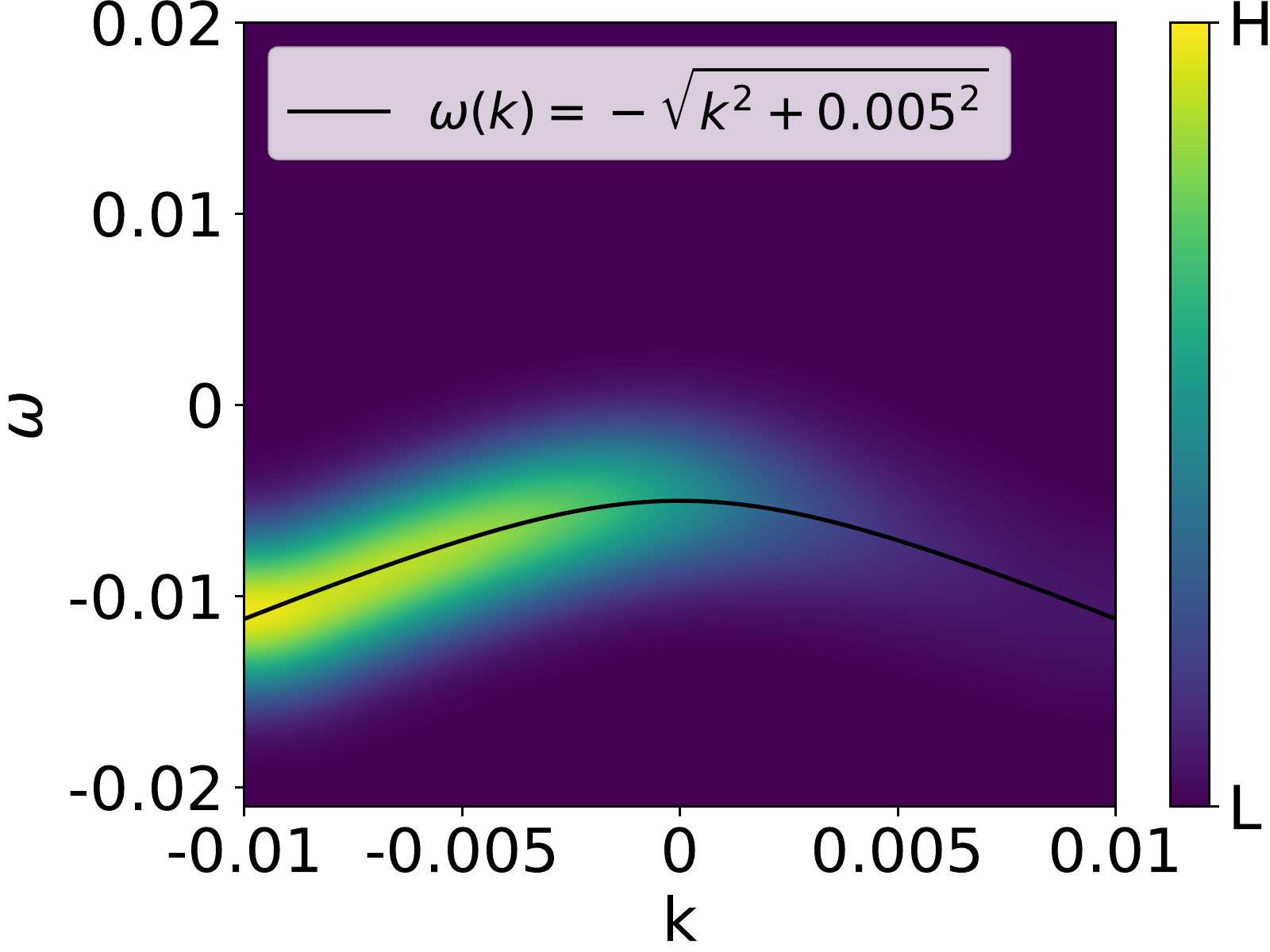}
    \includegraphics[scale=0.25]{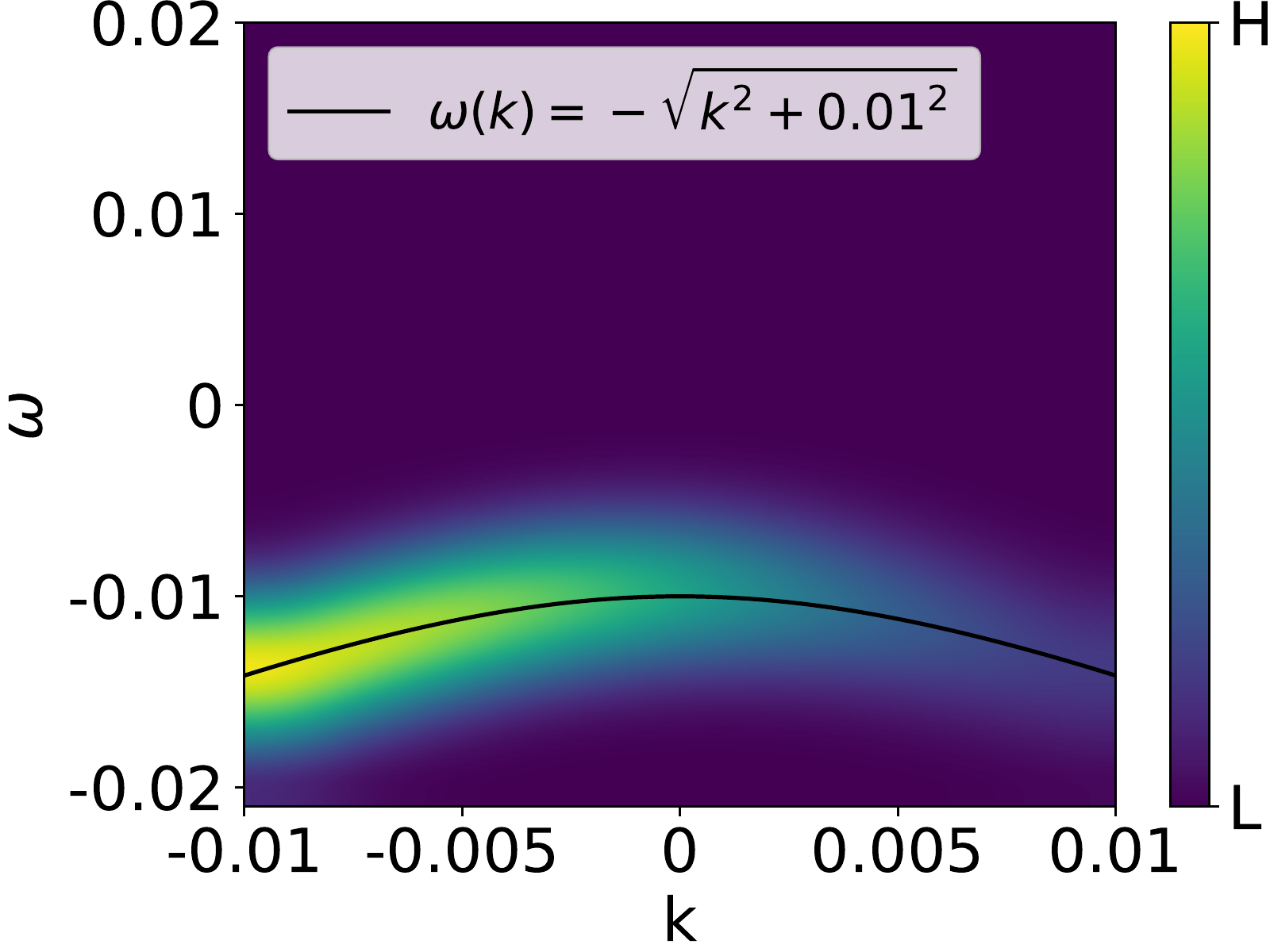}
    \caption{Equilibrium tr-ARPES calculations of a BCS superconducting system at equilibrium, with $\Delta=0.01$ (left panel) and $\Delta=0.01$ (right panel), $\epsilon_k=k$, $\gamma=0.0001$, $t_p=500$, $t=1000$, $\sigma=400$. The black curve shows $\omega(k)=-\sqrt{\epsilon_k^2+\Delta^2}$.}
    \label{fig:bcs}
  \end{center}
\end{figure}

Fig.~\ref{fig:bcs} shows two calculations of tr-ARPES signal of equilibrium BCS superconducting systems with different order parameters. Similar to the metal calculations, the tr-ARPES signal shown here follows $A(k,\omega)f(\omega)$ as expected, up to a broadening of the energy resolution due to finite width of the probe pulse.

\subsection{Nonequilibrium superconducting systems}\label{noneq}

Last but not least, we present non-equilibrium predictions of tr-ARPES signals with different $\Delta(t)$ profiles as input parameter. We fix the energy resolution by fixing width of the probe pulse, $\sigma=0.02\times2/\gamma$; we vary $t_p$ to probe the system at different times.

We consider the following non-equilibrium gap profiles:
\begin{itemize}
    \item constant order parameter with a quench
        \begin{equation}
            \Delta(t)=\Delta\theta(t)
            \label{eq:qgap}
        \end{equation}
    \item quenched order parameter with decay
        \begin{equation}
            \Delta(t)=\Delta\theta(t)e^{-t/T}
            \label{eq:qdecay}
        \end{equation}
    \item Gaussian order parameter
        \begin{equation}
            \Delta(t)=\left(\Delta e^{-\left(\frac{t-t_d}{T}\right)^2}-\Delta_0\right)\left[\theta(t)-\theta(t-2t_d)\right]
            \label{eq:gauss}
        \end{equation}
        where $\Delta_0$ is chosen such that $\Delta(t)=0$ when $t=0$ and $t=2t_d$.
\end{itemize}

\begin{figure}
  \begin{center}
    \includegraphics[scale=0.45]{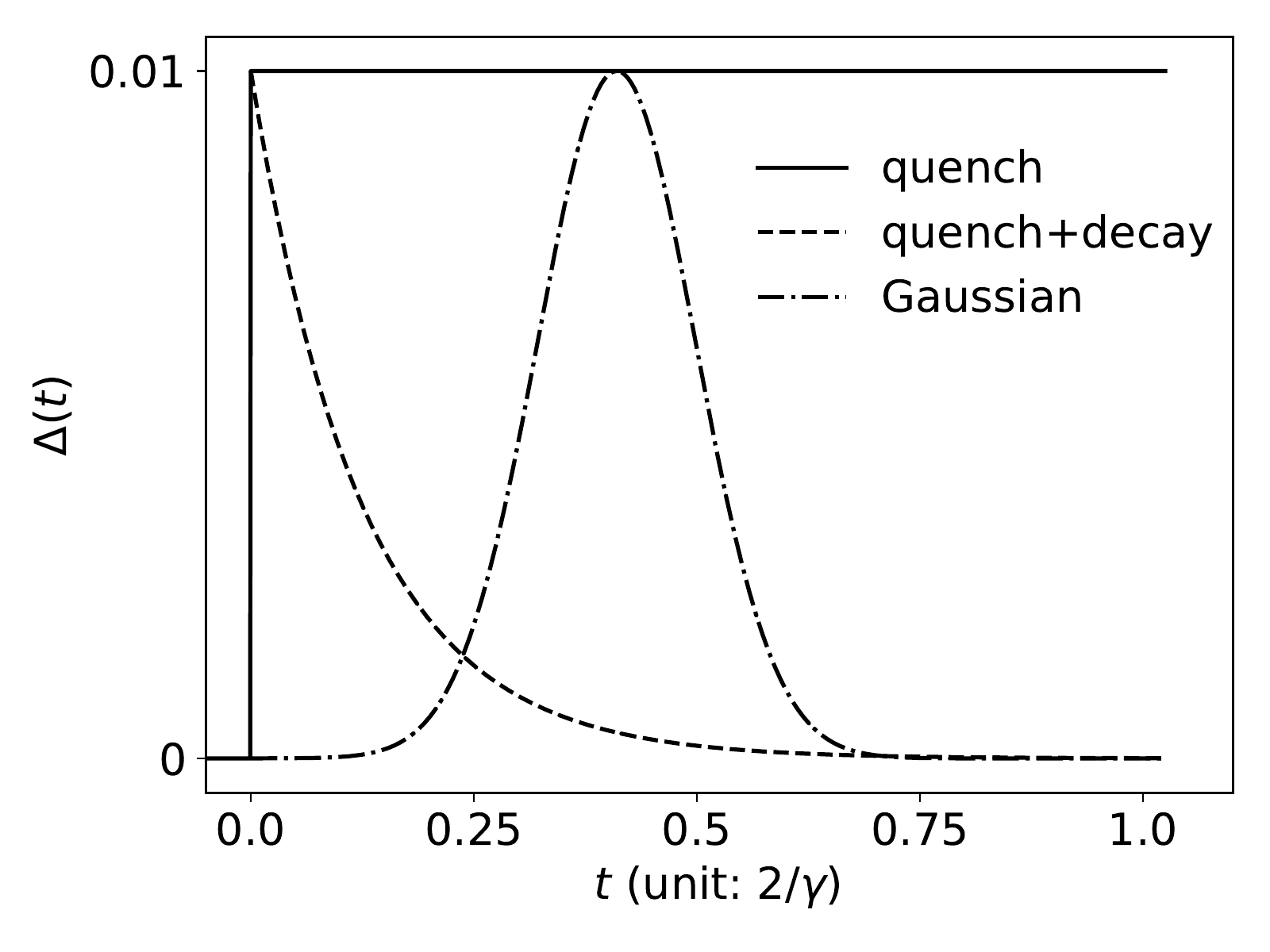}
    \caption{Schematic plot of various gap profiles considered in this work. }
    \label{fig:neqgaps}
  \end{center}
\end{figure}

A schematic plot of various $\Delta(t)$'s are shown in \fig{neqgaps}. Note that we always turn on non-equilibrium $\Delta(t)$ at $t=0$, and always consider a system initially at metallic state, i.e.: $\Delta(t<0)\equiv 0$. 

\subsubsection{Constant order parameter with a quench}

\begin{figure}[!htb]
    \centering
    \includegraphics[scale=0.7]{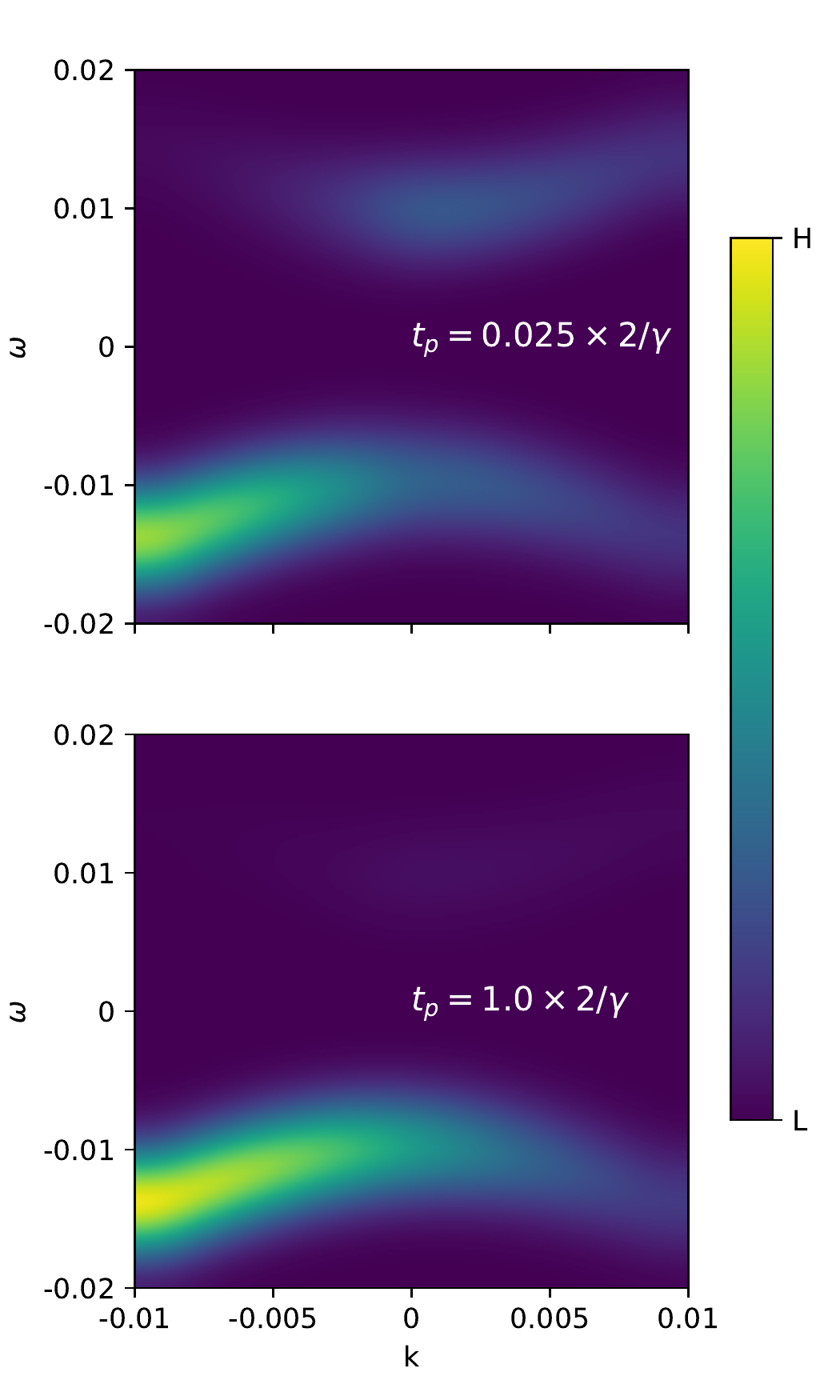}
    \caption{Non-equilibrium tr-ARPES calculations with BCS order parameter featuring constant with a quench, as mentioned in \eq{qgap}, with $\Delta=0.01$. The calculation is with $\epsilon_k=k$ and $\gamma=0.0001$. The width of the probe pulse is $\sigma=0.02\times2/\gamma$. The two different $t_p$'s, short and long time after quench, are indicated in the figures.}
    \label{fig:quench}
\end{figure}

First, we consider the system with a quenched order parameter, as described by \eq{qgap}. This situation can be viewed as a sudden opening of a superconducting gap at time $t=0$, before which the system is metallic. In this situation, we can rewrite $H_k$ as:
\begin{equation}
  H_k(t)=[1-\theta(t)]H_{k,metal}+\theta(t)H_{k,BCS},
\end{equation}
where $H_{k,metal}$ takes the form of \eq{hk} with $\Delta(t)=0$, while $H_{k,BCS}$ with $\Delta(t)=\Delta$. The resulting tr-ARPES signal as a function of time is shown in \fig{quench}. We note that the upper panel of \fig{quench}, which shows the tr-ARPES signal at a time right after the quench, resembles \fig{quenchpk}. This indicates that the relaxation caused by $\gamma$ has not yet affected the system at such time scale. However, as time goes, the tr-ARPES signal at positive energies gets relaxed, and eventually, at time of $\sim2/\gamma$, the signal goes to that described by $H_{k,BCS}$, as shown in the lower panel of \fig{quench}.

We then examine the occupation, $n_k$, at different $k$ values shortly after the quench:
\begin{equation}
    n_k(t)=\braket{c_k^\dagger(t)c_k(t)}={\text{Im}}G^<_k(t,t).
\end{equation}
$n_k(t)$ is essentially what tr-ARPES measures with a $\delta-$function probe \cite{aoki14rmp}.

\begin{figure}
  \begin{center}
    \includegraphics[scale=0.45]{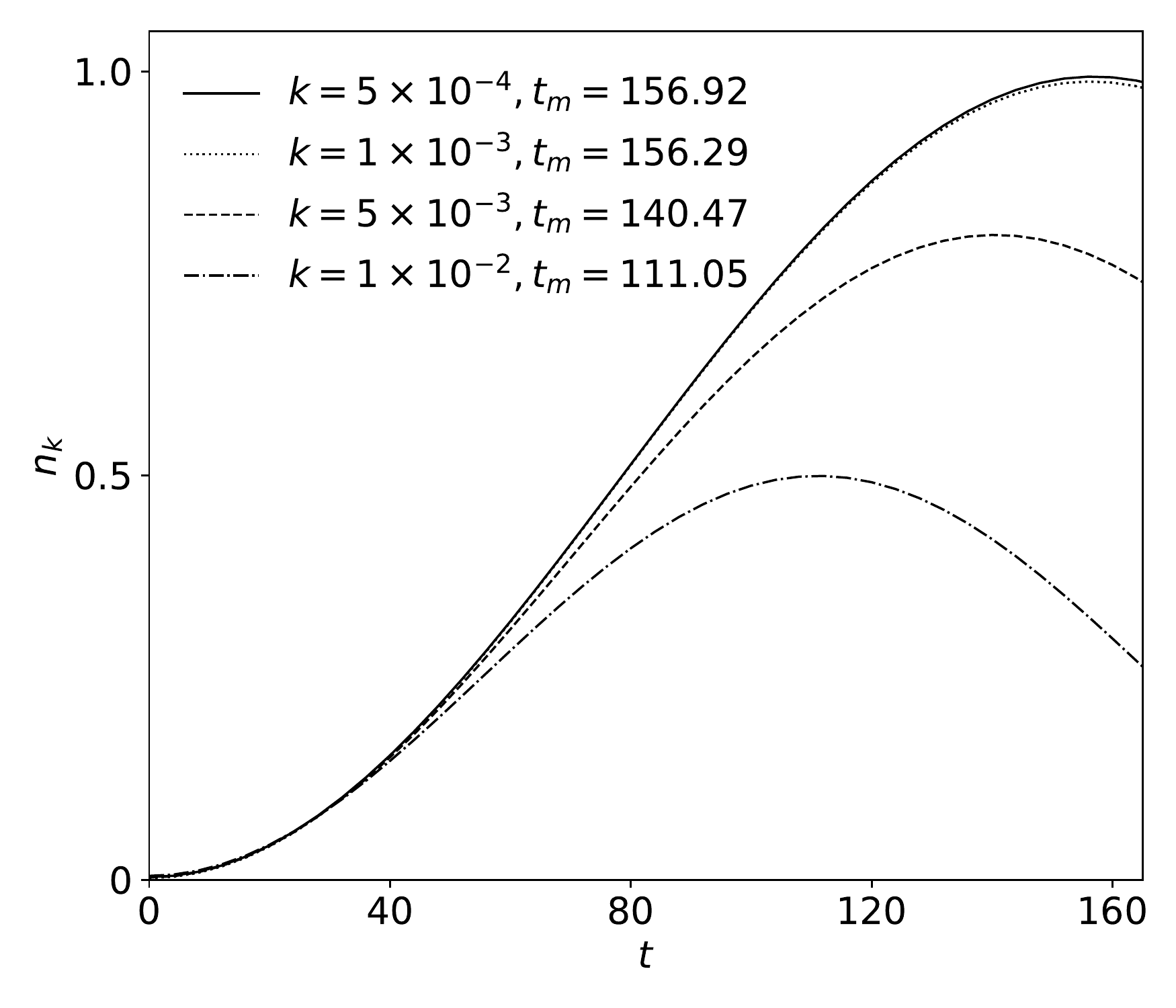}
    \caption{Occupation number, $n_k(t)$, of the quench system \eq{qgap} with $\Delta=0.01$. The times when $n_k$ peaks, $t_m$'s, are shown in the caption along with their corresponding $k$ values.}
    \label{fig:qnk}
  \end{center}
\end{figure}

When $t$ is small, i.e., right after quench, we should expect our system being more \emph{metallic}, namely, all what the quench does is projecting the pre-quench states onto post-quench basis. What we should expect for $k>0$ is that $n_k=0$ right after the quench, then $n_k$ peaks after some time scale that is defined by $1/E_k$. One should also expect $\lim_{k\to0^+}n_k\to 1$ at its peak value, since $n_k(t)$ shows Rabi oscillation around its equilibrium value, $1/2$ when $k=0$, in the post-quench system. Fig.~\ref{fig:qnk} shows $n_k(t)$ with various $k$'s and $\Delta=0.01$. We find that the $t$ and $k$ dependence of $n_k(t)$ are as expected, and $n_k(t)$ peaks at time $t_m=\pi/(2E_k)$, with $E_k$ defined in \eq{Euv}.

\subsubsection{Quenched order parameter with decay}

\begin{figure}[!htb]
    \centering
    \includegraphics[scale=0.43]{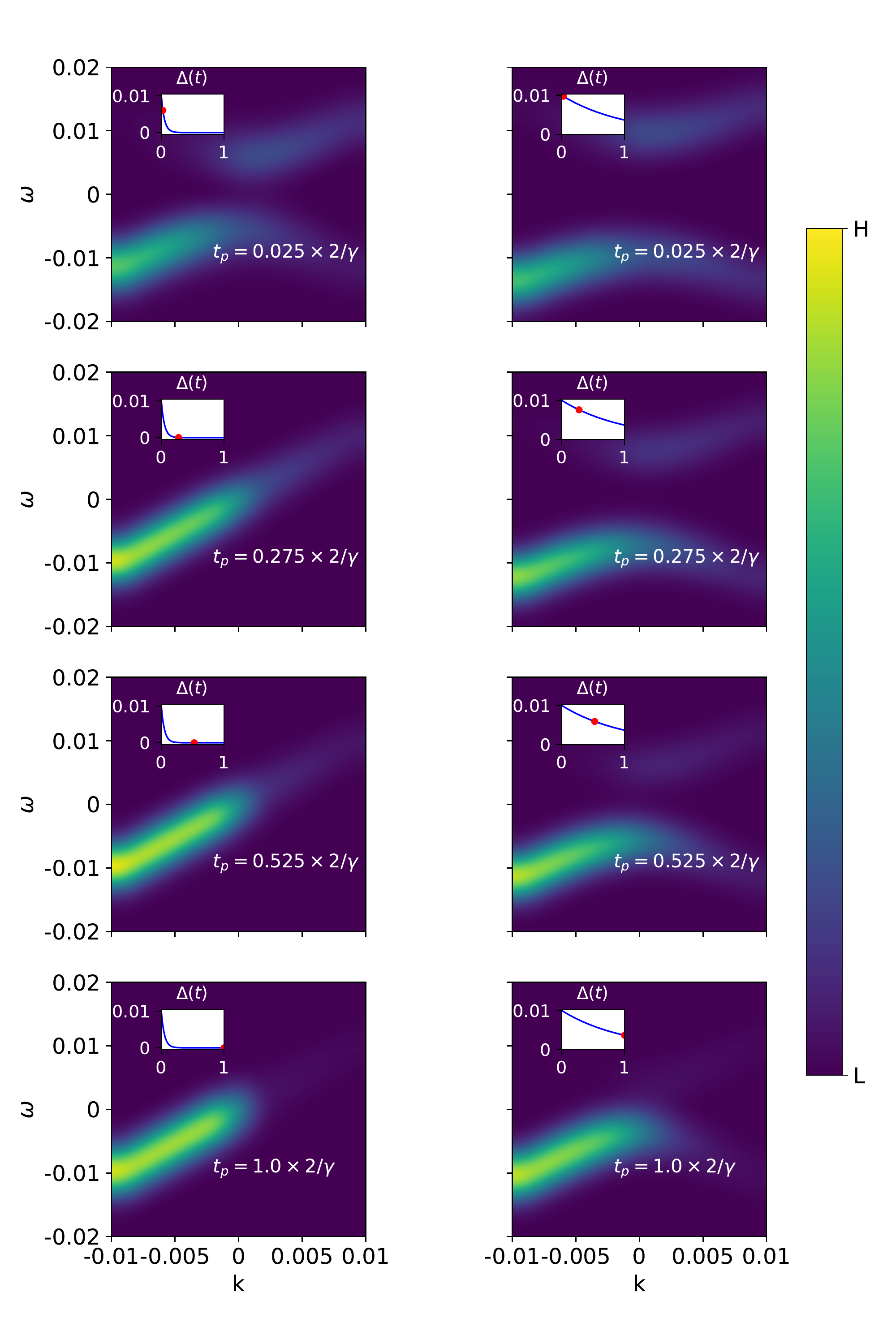}
    \caption{Non-equilibrium tr-ARPES calculations with BCS order parameters featuring quenched order parameters that decay, as mentioned in \eq{qdecay}, with $\Delta=0.01$, $T=0.05\times2/\gamma$ (left column) and $1\times2/\gamma$ (right column). The calculations are with $\epsilon_k=k$ and $\gamma=0.0001$. The width of the probe pulse is $\sigma=0.02\times2/\gamma$, and various $t_p$'s are indicated in the figures. The insets of the panels show gap profiles, $\Delta(t)$, as functions of time in unit of $2/\gamma$ in blue curves, and the instantaneous gap at $t_p$'s in red dots.}
    \label{fig:qdecaysl}
\end{figure}

Next we consider the the system with a quenched order parameter that decays right after the quench, as described by \eq{qdecay}. We consider two time scales of how fast the order parameter decays: $T  \ll 2/\gamma$ and $T=2/\gamma$. The resulting tr-ARPES signal as a function of time is shown in Fig.~\ref{fig:qdecaysl}. 

The left column of Fig.~\ref{fig:qdecaysl} shows tr-ARPES of the order parameter that decays with time scale $T \ll 2/\gamma$, while the right column, $T=2/\gamma$. Comparing left and right columns, one sees similar short-time behavior of tr-ARPES signals but different intermediate- and long-time behaviors. At short-time (the first panels of both columns), both small and large $T$ give signals similar to that of a quench, despite that the signals from small $T$ gives smaller gap than that from large $T$. For intermediate times (the second and third panels of both columns), the tr-ARPES signals of small and large $T$ show different behaviors: while the large $T$ signals still being gapped and quench-like, the small $T$ signals becomes gapless, and with a non-zero signal at positive $(k,\omega)$. Then, at long time, the tr-ARPES signal of small $T$ seems metallic while that of large $T$ being BCS-like -- corresponding qualitatively to their instantaneous BCS gaps.

\subsubsection{Gaussian order parameter}

\begin{figure}[!htb]
    \centering
    \includegraphics[scale=0.43]{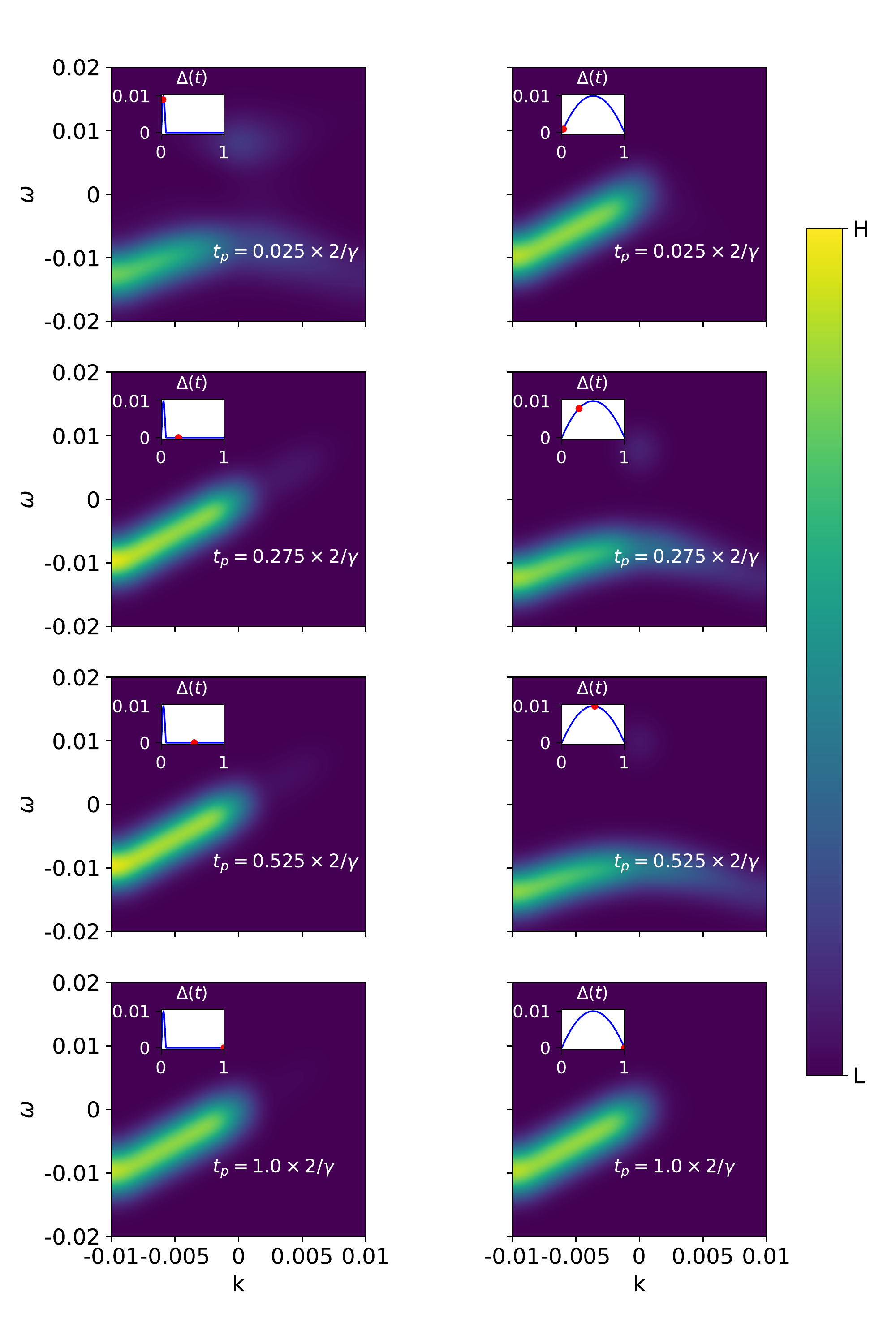}
    \caption{Non-equilibrium tr-ARPES calculations with BCS order parameters featuring Gaussian order parameters, as mentioned in \eq{gauss}, with $\Delta(t)_{max}=0.01$, $T=0.05\times2/\gamma$ (left column) and $3\times2/\gamma$ (right column), $t_d=0.035\times2/\gamma$ (left column) and $0.5\times2/\gamma$ (right column). The calculations are with $\epsilon_k=k$ and $\gamma=0.0001$. The width of the probe pulse is $\sigma=0.02\times2/\gamma$, and various $t_p$'s are indicated in the figures. The insets of the panels show gap profiles, $\Delta(t)$, as functions of time in unit of $2/\gamma$ in blue curves, and the instantaneous gap at $t_p$'s in red dots.}
    \label{fig:gausssl}
\end{figure}

Finally, we consider the the system with a Gaussian order parameter, as described by \eq{gauss}. The resulting tr-ARPES signal is shown in \fig{gausssl}.

The left column of \fig{gausssl} shows tr-ARPES signals of a Gaussian order parameter with $T=0.05\times2/\gamma$ and $t_d=0.035\times2/\gamma$, while the right column shows that with $T=3\times2/\gamma$ and $t_d=0.5\times2/\gamma$. At first glance, the short- and intermediate- time behaviors of tr-ARPES signals (first to third panels of both columns) seem quite different, while the long-time (last panels of both columns) share similarities. However, if we compare the tr-ARPES signals to the instantaneous order parameters at $t_p$, both columns show tr-ARPES signals that almost follow the instantaneous order parameters, except for a slight quench-like peak near $k_F$ for the first panel of the left column and the second and third panels of the right column, where the instantaneous order parameters 
approach the maximum values.
In addition, the quench-like peak is more apparent in the left column than the right column, which may come from that in the left column, the increment of the order parameter is faster than the right column, causing a more significant quench-like peak at low $k$. Another feature to notice is that, this quench-like peak only happens near $k_F$, which is different from an actual quench peak that also extends to $k>k_F$. This may be related to a slower increment of the Gaussian order parameter than that of a sudden quench.

\section{Discussion} \label{discuss}

The importance of equilibrium ARPES measurements of the electron spectral function to our understanding of unconventional superconductors is clear.  To bring tr-ARPES to a similar level of scientific impact, it is necessary to be able to draw conclusions from data about the underlying physics even though what tr-ARPES measures is considerably more challenging to interpret. It is already known from pump-probe studies of optical conductivity and other properties that superconductors support rich non-equilibrium behavior including significant differences between conventional and unconventional superconductors.

The approach developed here is intended to improve the practical ability of tr-ARPES to probe the non-equilibrium fermionic properties of a complex material. It is worth noting that an implicit assumption we made here is that the pump pulse only affects the system by changing the BCS superconducting gap.
However, in actual experiments, the pump field may modify the electron distribution in other ways, where one needs to consider the modified post-pump initial state that changes the boundary condition of the lesser Green's function. This can be described by explicitly solving the time dependent Schr\"odinger equation with light-matter interaction\cite{devereaux13prx,devereaux15prb,sentef16prb,demlersc17,murakami17,devereaux17}, although that will be more time-consuming. 
An alternative and more efficient way to approximate this is effectively raising the electron temperature, $T_e$ \cite{allen87prl}, which is incorporated by increasing the temperature of $\Sigma^<(t,t')$ (see Appendix \ref{app:temperature} for details).

Some obvious extensions to the theory presented here are to non-s-wave superconductors and to incorporate some level of disorder (such as Refs. \onlinecite{mitra17prb,mitra18prl,mitra18arxiv}), as well as using self-consistently determined BCS order parameters (such as Refs. \onlinecite{spivak04prl, millis17nat,foster17prb}) as input to our theory. More challenging extensions would be to approach unconventional superconductivity more microscopically (e.g., via various proposed effective interactions or coupling to order parameters such as nematic or magnetic order) and to compute the superconducting properties of such systems self-consistently, as well as taking other microscopic effect of the pump into consideration, such as stripe melting \cite{fausti11sci,forst14prl} and possible phonon squeezing in $K_3C_{60}$ \cite{cavalleri16}.

But already our non-equilibrium calculations begin to show how various tr-ARPES signals indicate the qualitative differences among order parameter profiles. For example, our calculations show that, a signal at positive $(k,\omega)$ indicates that there may be a non-adiabatic change of BCS order parameter (e.g., quenched order parameter); while on the other hand, a time-varying tr-ARPES signal without such peaks may indicate an adiabatic change of BCS order parameter. Other possible ways of using these simulated results include computing and analyzing $\int dk\int_{\omega>0}d\omega~I(k,\omega,t)$, and modifying $s(t)$ in \eq{Ikwt} to see how underlying order parameter dynamics may be revealed by different probe pulses. We hope that the model developed here will be useful in the extraction of physics from experimental data, thus indicating constraints for more microscopic theories of superconductivity.

\section*{Acknowledgments}
We would like to thank A. Millis for valuable discussions. The authors were supported by the U.S. Department of Energy (DOE), Office of Science, Basic Energy Sciences (BES), under Contract No. DE-AC02-05-CH11231 within the Ultrafast Materials Science Program (KC2203).  T.M. acknowledges support from the EPiQS initiative of the Gordon and Betty Moore Foundation.


\section*{Appendix}
\appendix

\section{Derivation of equilibrium $G^<$'s from $\Sigma^<$}\label{app:eqgl}
In this section we present derivation of equilibrium $G^<$ from $\Sigma^<$ that we proposed in Section \ref{glesser}.

To calculate $G^<$ from \eq{glcal}, one first notice that, due to the causality of $G^R$ and $G^A$, the explicit integration limits are:
\begin{align}
  &G^<_k(t,t')\nonumber\\
  &=\frac{-\gamma}{2\pi}\int_{-\infty}^t dt_1\int_{-\infty}^{t'} dt_2\frac{G^R_k(t,t_1)G^A_k(t_2,t')}{t_1-t_2+i0^+}\nonumber\\
  &\qquad\qquad\qquad\times e^{-\gamma(t-t_1+t'-t_2)/2}.
  \label{eq:glcal2}
\end{align}

Then, we do the following change of variables:
\begin{align}
    T&=\frac{t_1+t_2}{2}\nonumber\\
    \tau&=t_1-t_2,
    \label{eq:trot}
\end{align}
and thus \eq{glcal2} becomes
\begin{align}
  &G^<_k(t,t')\nonumber\\
  &=\frac{-\gamma}{2\pi}e^{-\gamma(t+t')/2}\nonumber\\
  &\times\int_{-\infty}^{\frac{t+t'}{2}} dT\int_{-\infty}^\infty d\tau\frac{G^R_k(t,t_1)G^A_k(t_2,t')}{\tau+i0^+}e^{\gamma T},
  \label{eq:glcal3}
\end{align}
where $t_1$ and $t_2$ are functions of $T$ and $\tau$ as in \eq{trot}.

Now, let us first plug in $G^{R/A}$ of metal, namely
\begin{align}
    G^R(t,t_1)&=-i\theta(t-t_1)e^{-i\epsilon_k(t-t_1)}\nonumber\\
    G^A(t_2,t')&=i\theta(t'-t_2)e^{i\epsilon_k(t'-t_2)},
\end{align}
then we get $G^<$ of metal from \eq{glcal3}
\begin{align}
  &G^<_k(t,t')\nonumber\\
  &=\frac{-\gamma}{2\pi}e^{-\gamma(t+t')/2}e^{-i\epsilon_k(t-t')}\int_{-\infty}^{\frac{t+t'}{2}} e^{\gamma T} dT\int_{-\infty}^\infty \frac{e^{i\epsilon_k\tau}}{\tau+i0^+}d\tau\nonumber\\
  &=i\theta(-\epsilon_k)e^{-i\epsilon_k(t-t')}.
  \label{eq:glcalmetal}
\end{align}
At zero temperature, we have $\theta(-\epsilon_k)=f(\epsilon_k)$, thus giving us exactly the $G^<$ one gets via \eq{grakldef}.

The $G^<$ calculation for BCS superconductor is slightly more complicated, as when one works in Nambu-spinor basis, $G^{R/A}$ becomes a 2$\times$2 matrix 
\begin{widetext}
  \begin{align}
    G^R_k(t,t')&=\hp{-i[u_k^2e^{-iE_k(t-t')}+v_k^2e^{iE_k(t-t')}]}{iu_kv_k[e^{iE_k(t-t')}-e^{-iE_k(t-t')}]}{iu_kv_k[e^{iE_k(t-t')}-e^{-iE_k(t-t')}]}{-i[u_k^2e^{iE_k(t-t')}+v_k^2e^{-iE_k(t-t')}]}\nonumber\\
    G^A_k(t,t')&=[G^R(t',t)]^\dagger\nonumber\\
    &=
    \hp{i[u_k^2e^{iE_k(t'-t)}+v_k^2e^{-iE_k(t'-t)}]}{-iu_kv_k[e^{-iE_k(t'-t)}-e^{iE_k(t'-t)}]}{-iu_kv_k[e^{-iE_k(t'-t)}-e^{iE_k(t'-t)}]}{i[u_k^2e^{-iE_k(t'-t)}+v_k^2e^{iE_k(t'-t)}]},
    \label{eq:grabcs}
  \end{align}
\end{widetext}
with \cite{schriefferscbook}
\begin{equation}
  E_k=\sqrt{\epsilon_k^2+\Delta^2},\ u_k^2=\frac{1}{2}(1+\frac{\epsilon_k}{E_k}),\ v_k^2=1-u_k^2,
  \label{eq:Euv}
\end{equation}

Plugging these into \eq{glcal3}, and do time-rotation as in \eq{trot}, we get
\begin{align}
  &G^<_k(t,t')\nonumber\\
  &=\frac{-\gamma}{2\pi}e^{-\gamma(t+t')/2}\nonumber\\
  &\times\int_{-\infty}^{\frac{t+t'}{2}} dT\int_{-\infty}^\infty d\tau\frac{u_k^2e^{-iE_k(t-t'-\tau)}+v_k^2e^{iE_k(t-t'-\tau)}}{\tau+i0^+}e^{\gamma T}\nonumber\\
  &=i[u_k^2\theta(-E_k)e^{-iE_k(t-t')}+v_k^2\theta(E_k)e^{iE_k(t-t')}]\nonumber\\
  &=iv_k^2e^{iE_k(t-t')},
  \label{eq:glcalbcs}
\end{align}
which is exactly the $G^<$ one gets from direct calculation (\eq{grakldef}). Note that the last equality comes from the fact that $E_k$ is always positive due to its definition in \eq{Euv}.

\section{The non-dissipative nature of \eq{glnodis}, and the $\gamma\to 0$ limit of \eq{glcal}} \label{app:nondis}
To simulate tr-ARPES signal of a quenched system, i.e. \eq{qhami}, using \eq{glnodis}, one uses $G^R_k$ and $G^A_k$ of the BCS superconducting system that is described by $H_{BCS}$, i.e. \eq{grabcs}, while taking $n_{0,k}$ to be that of a metal. Therefore, one gets the quenched lesser Green's function
\begin{align}
  &G^<_{k,quench}(t,t')\nonumber\\
  &=iG^R_{k,BCS}(t,t_0)\hp{f(\epsilon_k)}{0}{0}{1-f(\epsilon_{-k})}G^A_{k,BCS}(t_0,t'),
  \label{eq:glqnodiss}
\end{align}
Plugging $G^{R/A}_{k,BCS}$ into the above equation and only taking the normal component, one gets
\begin{align}
    &G^<_{k,quench}(t,t')\nonumber\\
    &=if(\epsilon_k)\bigg[u_k^4e^{-iE_k(t-t')}+v_k^4e^{iE_k(t-t')}\nonumber\\
    &\quad\quad+u_k^2v_k^2e^{-iE_k(t+t'-2t_0)}+u_k^2v_k^2e^{iE_k(t+t'-2t_0)}\bigg]\nonumber\\
    &\quad+i\big[1-f(\epsilon_{-k})\big]u_k^2v_k^2\bigg[e^{-iE_k(t-t')}+e^{iE_k(t-t')}\nonumber\\
    &\quad\quad\quad-e^{-iE_k(t+t'-2t_0)}-e^{iE_k(t+t'-2t_0)}\bigg].
    \label{eq:glqnodiss2}
\end{align}
When computing tr-ARPES signal with \eq{glqnodiss2}, a positive energy peak shows up due to a non-zero value of $u_k^2v_k^2$. The absence of dissipation prevents such excitation from relaxing.

One can also obtain \eq{glqnodiss} from \eq{glcal} by taking $\gamma\to 0$ limit. When $\gamma$ is small and $t, t'$ are positive but small, one has:
\begin{align}
    \int_{-\infty}^t dt_1\int_{-\infty}^{t'} dt_2&\simeq\int_{-T}^t dt_1\int_{-T}^{t'} dt_2\nonumber\\
    &\simeq\int_{-T}^0 dt_1\int_{-T}^0 dt_2\nonumber\\
    &\simeq\int_{-\infty}^0 dt_1\int_{-\infty}^0 dt_2,
\end{align}
where $T\sim 1/\gamma$ is the effective time scale caused by level broadening of $G^{R/A}$. Then one can approximate $G^<$ with:
\begin{align}
    &G^<_k(t,t')\nonumber\\
    &\simeq\frac{-\gamma}{2\pi}\int_{-\infty}^0 dt_1\int_{-\infty}^{0} dt_2\frac{G^R_k(t,t_1)G^A_k(t_2,t')}{t_1-t_2+i0^+}\nonumber\\
    &\qquad\qquad\qquad\times e^{-\gamma(t-t_1+t'-t_2)/2}.
  \label{eq:glsg}
\end{align}
Now, $G^R_k(t,t')$ and $G^A_k(t,t')$ for the quench problem can be derived analytically by solving the partial differential equation
\begin{equation}
    i\partial_tG^R_k(t,t')=H_k(t)G^R_k(t,t'),
    \label{eq:grpde}
\end{equation}
with 
\begin{equation}
    H_k(t)=\theta(-t)H_{k,metal}+\theta(t)H_{k,BCS},
\end{equation}
where
\begin{align}
    H_{k,metal}&=\hp{\epsilon_k}{0}{0}{-\epsilon_k}\nonumber\\
    H_{k,BCS}&=\hp{\epsilon_k}{\Delta}{\Delta}{-\epsilon_k},
\end{align}
with initial condition $G^R(t',t')=-i$. Let us write $H_{k,metal}=H_{k,m}$ and $H_{k,BCS}=H_{k,Sc}$.

The solution to \eq{grpde} is
\begin{equation}
    G^R_k(t,t')=
    \begin{cases}
    U_1(t,t')(-i) & t'<t<0\\ 
    U_2(t,0)U_1(0,t')(-i) & t'<0<t\\
    U_2(t,t')(-i) & 0<t'<t,
    \end{cases}
\end{equation}
where 
\begin{align}
    U_{1/2}(t,t')&=T\left[exp(-i\int_{t'}^tH_{k,m/Sc}ds)\right]\nonumber\\
    &=exp\left[-iH_{k,m/Sc}(t-t')\right].
\end{align}
Then, \eq{glsg} becomes:
\begin{widetext}
\begin{align}
    G^<_k(t,t')&\simeq-\frac{\gamma}{2\pi}\int_{-\infty}^0 dt_1\int_{-\infty}^{0} dt_2\frac{U_2(t,0)U_1(0,t_1)(-i)iU^\dagger_1(0,t_2)U^\dagger_2(t',0)}{t_1-t_2+i0^+}e^{-\gamma(t-t_1+t'-t_2)/2}\nonumber\\
    &=\frac{-\gamma}{2\pi}e^{-\gamma(t+t')/2}U_2(t,0)\int_{-\infty}^0 dt_1\int_{-\infty}^{0} dt_2\frac{e^{iH_{k,m}(t_1-t_2)} e^{\gamma(t_1+t_2)/2}}{t_1-t_2+i0^+}U^\dagger_2(t',0)\nonumber\\
    &=ie^{-\gamma(t+t')/2}U_2(t,0)\hp{f(\epsilon_k)}{0}{0}{f(-\epsilon_k)}U^\dagger_2(t',0)\nonumber\\
    &=iG^R_{k,BCS}(t,t_0)\hp{f(\epsilon_k)}{0}{0}{f(-\epsilon_k)}G^A_{k,BCS}(t_0,t'),
\end{align}
\end{widetext}
with $t_0=0$. The last equality holds when $\gamma\to 0$ and $t, t'$ sufficiently small. Also notice that $f(-\epsilon_k)=1-f(\epsilon_{k})=1-f(\epsilon_{-k})$ for system with time-translational invariance.

Hence we have recovered \eq{glqnodiss}, the non-dissipating limit of the lesser Green's function of a quench BCS system.

\section{Tr-ARPES signals with thermal electron distribution} \label{app:temperature}
The lesser Green's function, with raised temperature, can be calculated using \eq{glkeldysh}, with 
\begin{align}
    \Sigma^<(t_1,t_2,T_e)&=i\gamma\int\frac{d\omega}{2\pi}f(\omega,T_e)e^{-i\omega(t_1-t_2)},
  \label{eq:slapproxT}
\end{align}
where $f(\omega,T_e)$ is the Fermi-Dirac distribution at temperature $T_e$. Here we assume that $T_e$ changes much slower than the time scale set by $\gamma$, so that this thermal distribution of bath electrons can be considered to be static. This assumption can, in principle, be relaxed by modifying $f(\omega,T_e)$.

\begin{figure}
    \centering
    \includegraphics[scale=0.4]{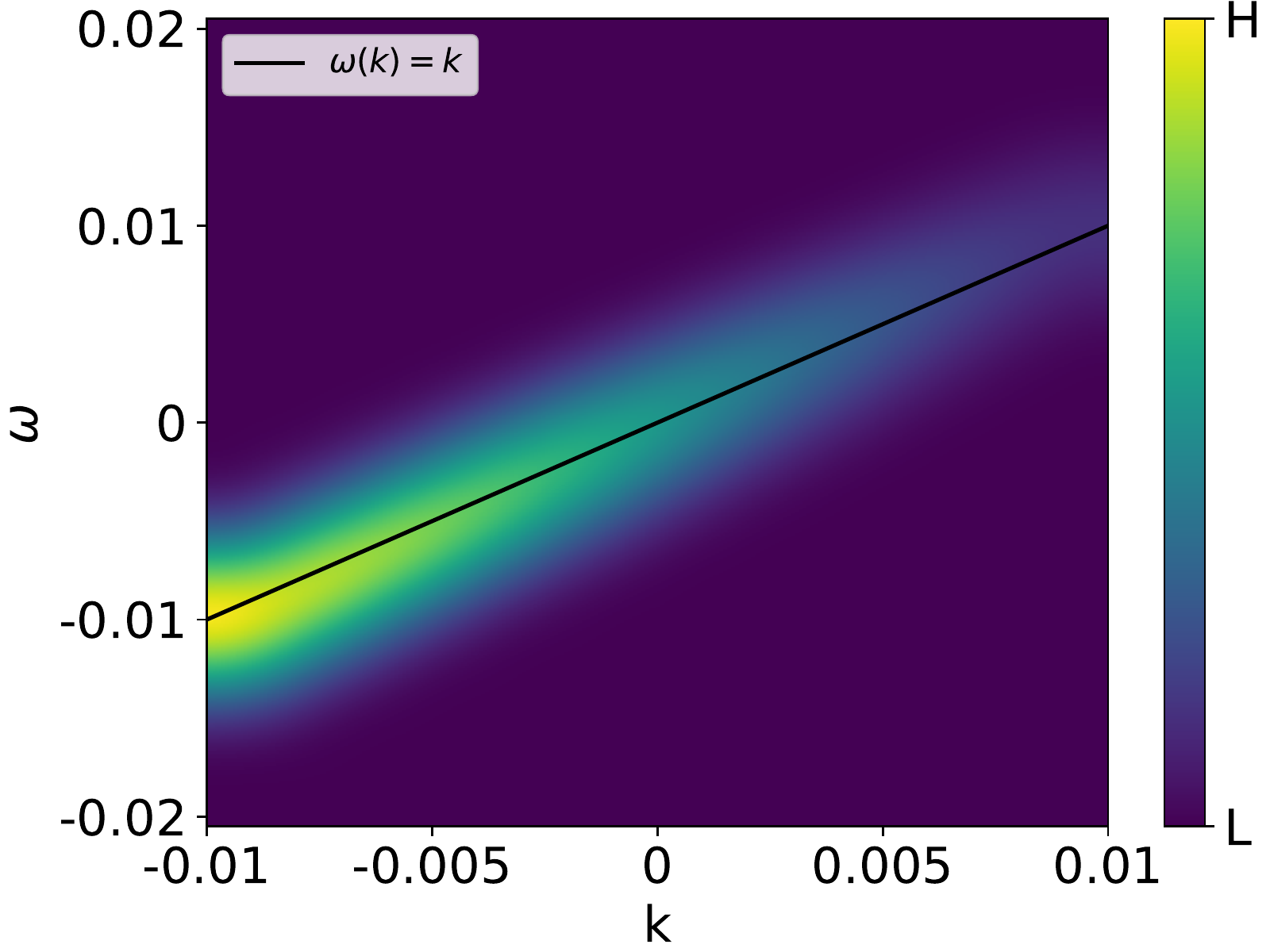}
    \includegraphics[scale=0.25]{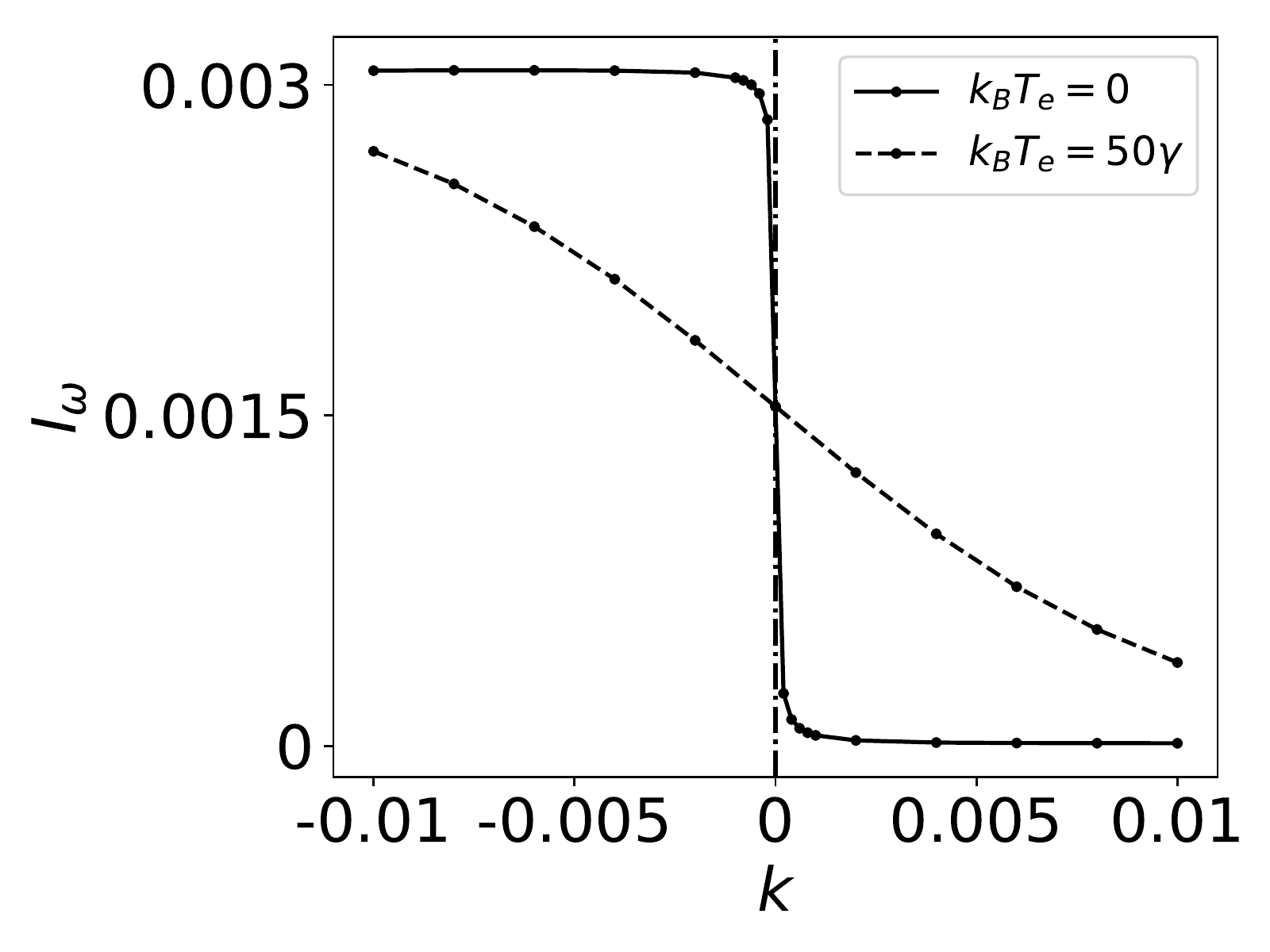}
    \includegraphics[scale=0.25]{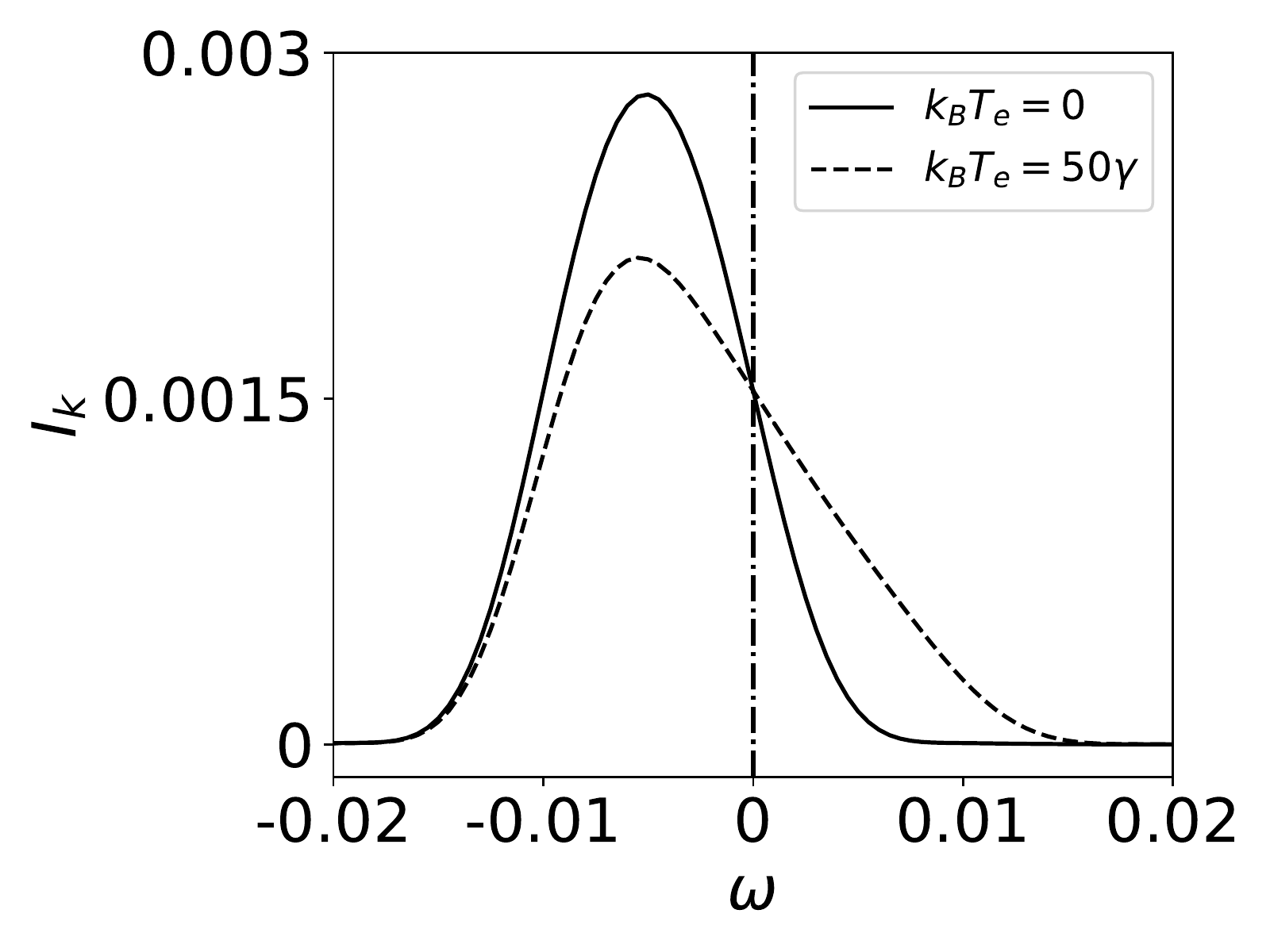}
    \caption{Equilibrium calculations of metal at non-zero temperature: $k_BT_e=50\gamma$. Upper panel: tr-ARPES signal with $\epsilon_k=k$, $\gamma=0.0001$, $t_p=500$, $t=1000$, $\sigma=400$. The black curve shows $\omega(k)=\epsilon_k$. Lower panels: $I_\omega(k,t)$ (left) and $I_k(\omega,t)$ (right) at $k_BT_e=0$ (solid line with dotted data points) and $k_BT_e=50\gamma$ (dashed line with dotted data points). The vertical dash-dotted lines show where $k$ (left) or $\omega$ (right) is zero.}
    \label{fig:metalT}
\end{figure}

\begin{figure}
    \centering
    \includegraphics[scale=0.4]{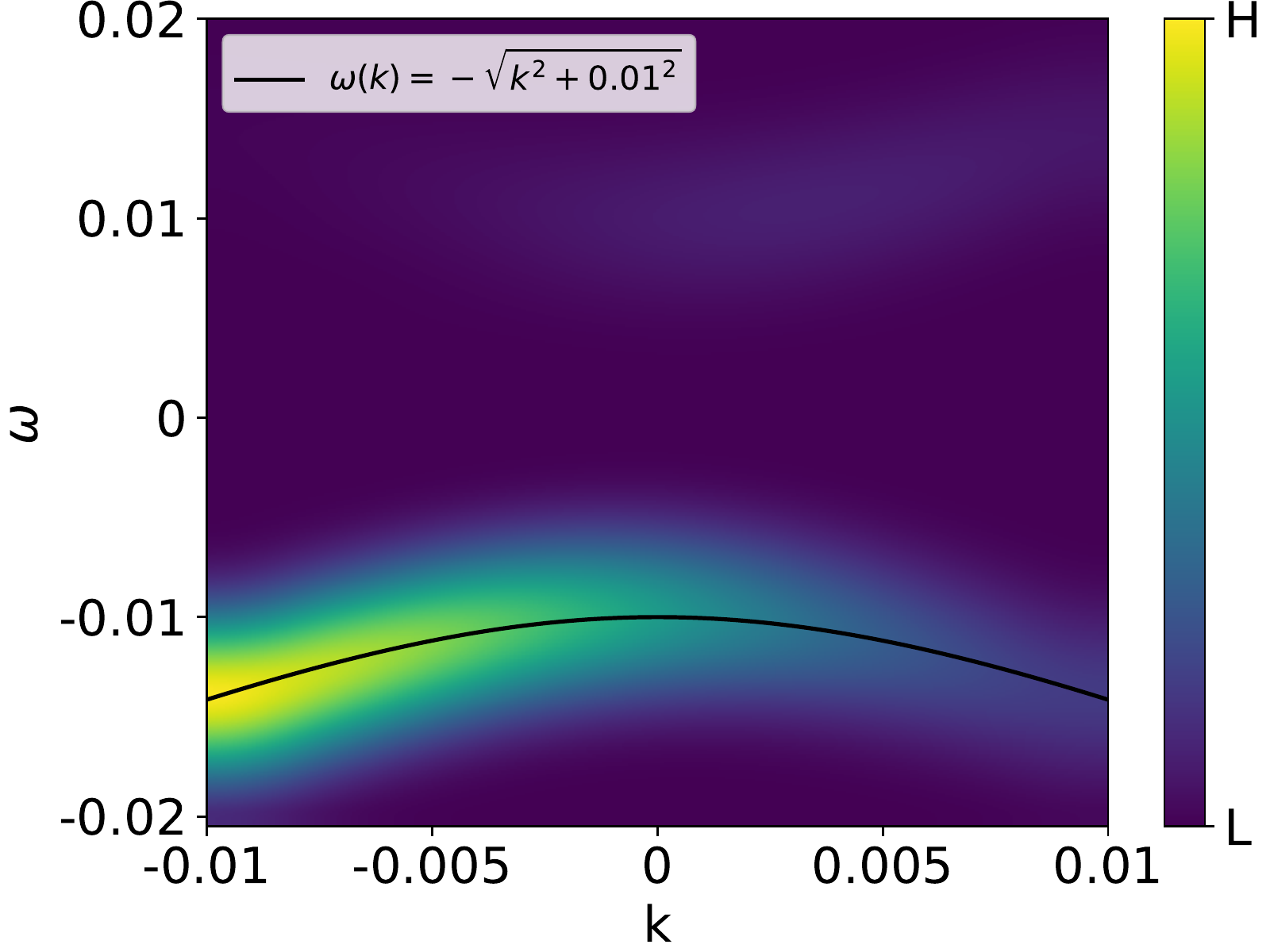}
    \includegraphics[scale=0.25]{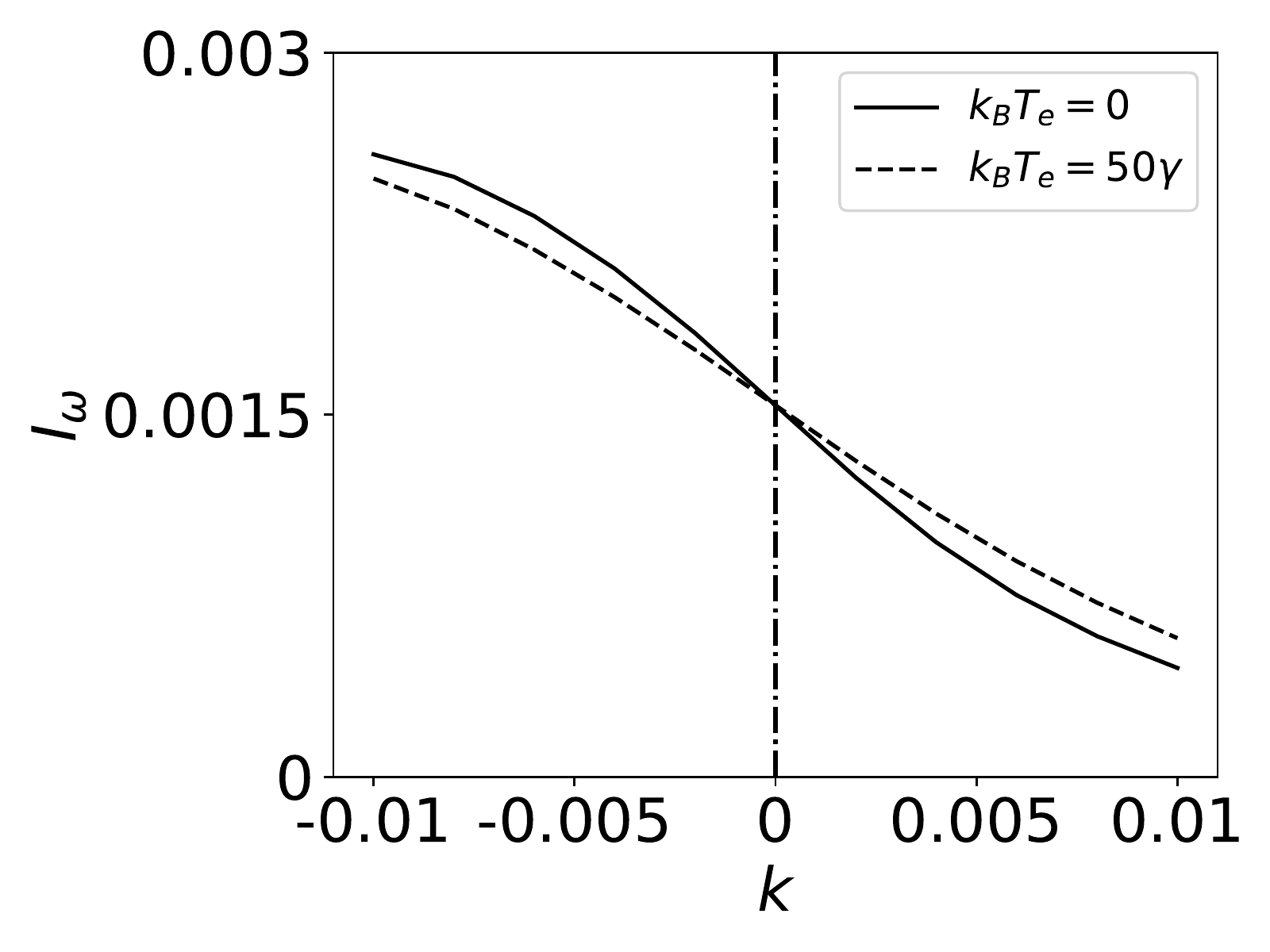}
    \includegraphics[scale=0.25]{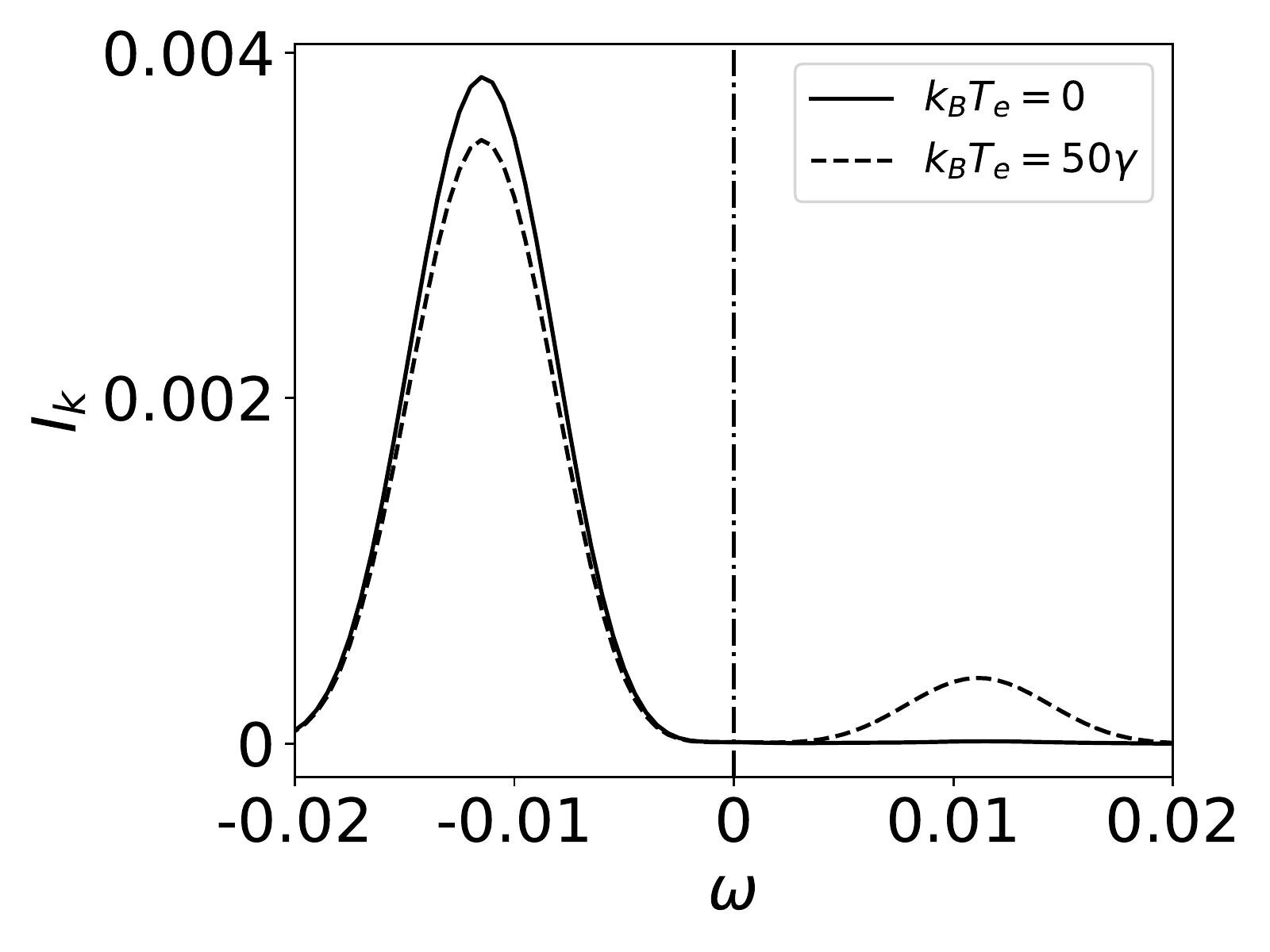}
    \caption{Equilibrium calculations of a BCS superconducting system with $\Delta=0.01$ at non-zero temperature: $k_BT_e=50\gamma$. Upper panel: tr-ARPES signal with $\epsilon_k=k$, $\gamma=0.0001$, $t_p=500$, $t=1000$, $\sigma=400$. The black curve shows $\omega(k)=-\sqrt{\epsilon_k^2+\Delta^2}$. Lower panels: $I_\omega(k,t)$ (left) and $I_k(\omega,t)$ (right) at $k_BT_e=0$ (solid line) and $k_BT_e=50\gamma$ (dashed line). The vertical dash-dotted lines show where $k$ (left) or $\omega$ (right) is zero.}
    \label{fig:BCST}
\end{figure}

\begin{figure}
    \centering
    \includegraphics[scale=0.7]{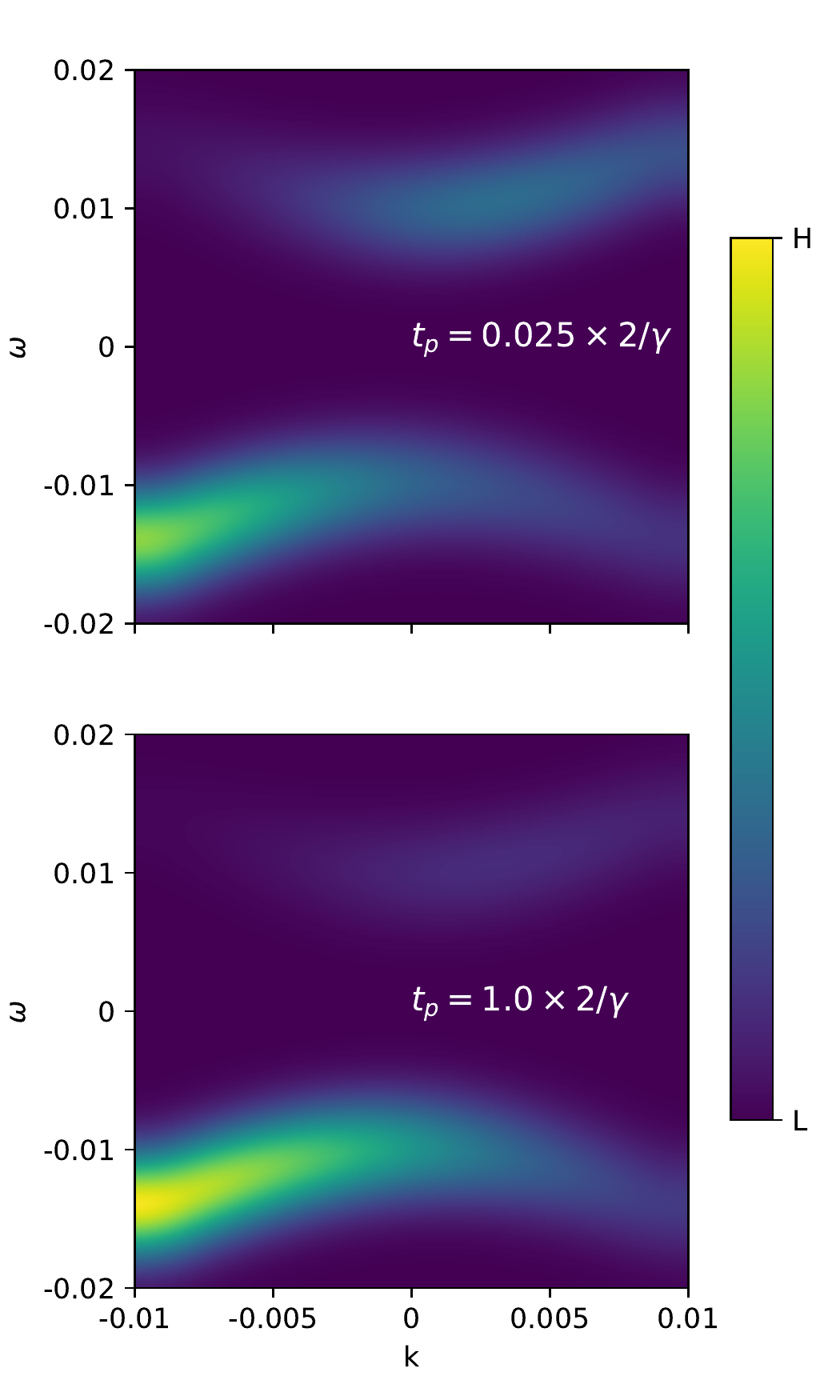}
    \includegraphics[scale=0.25]{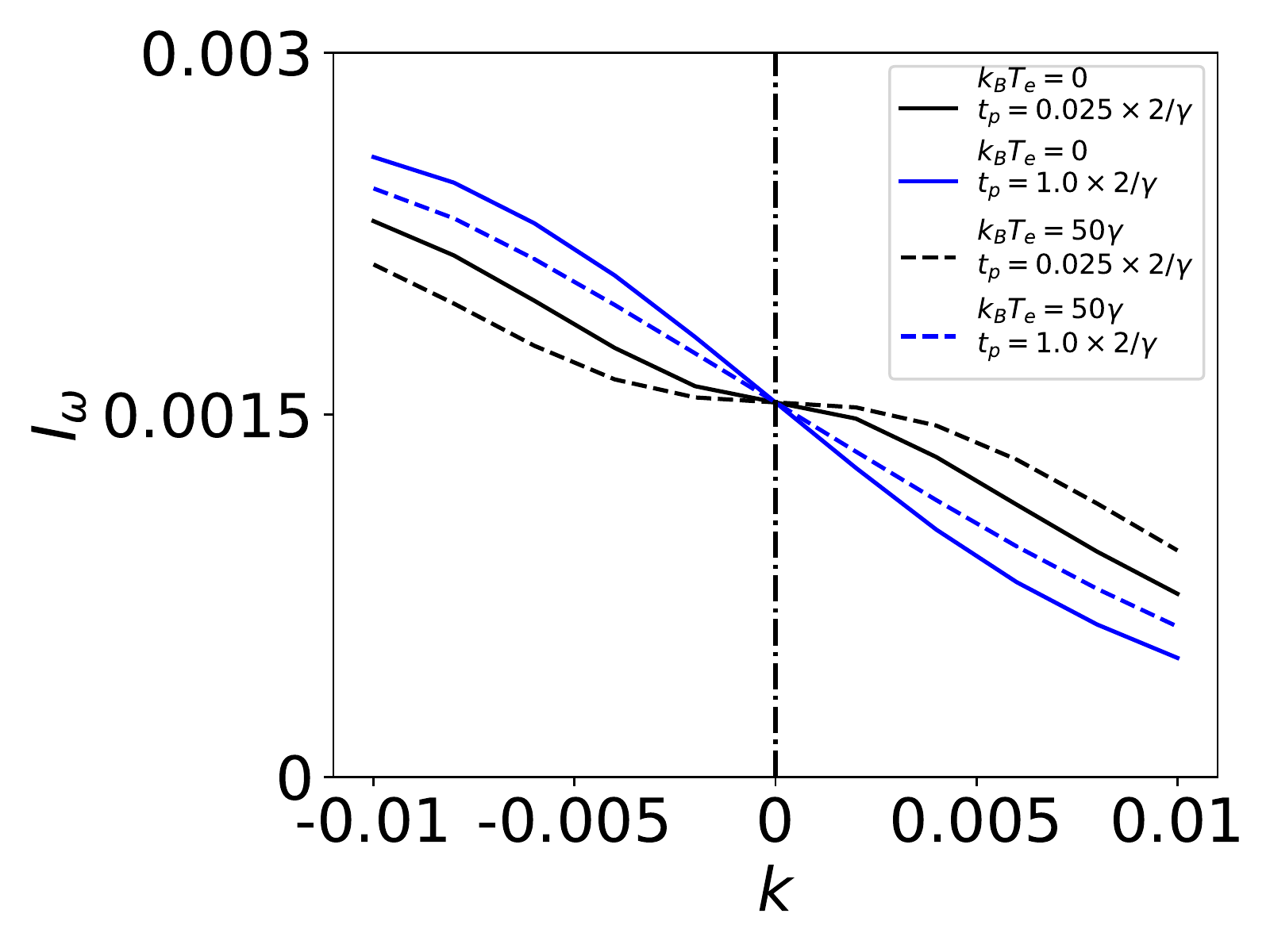}
    \includegraphics[scale=0.25]{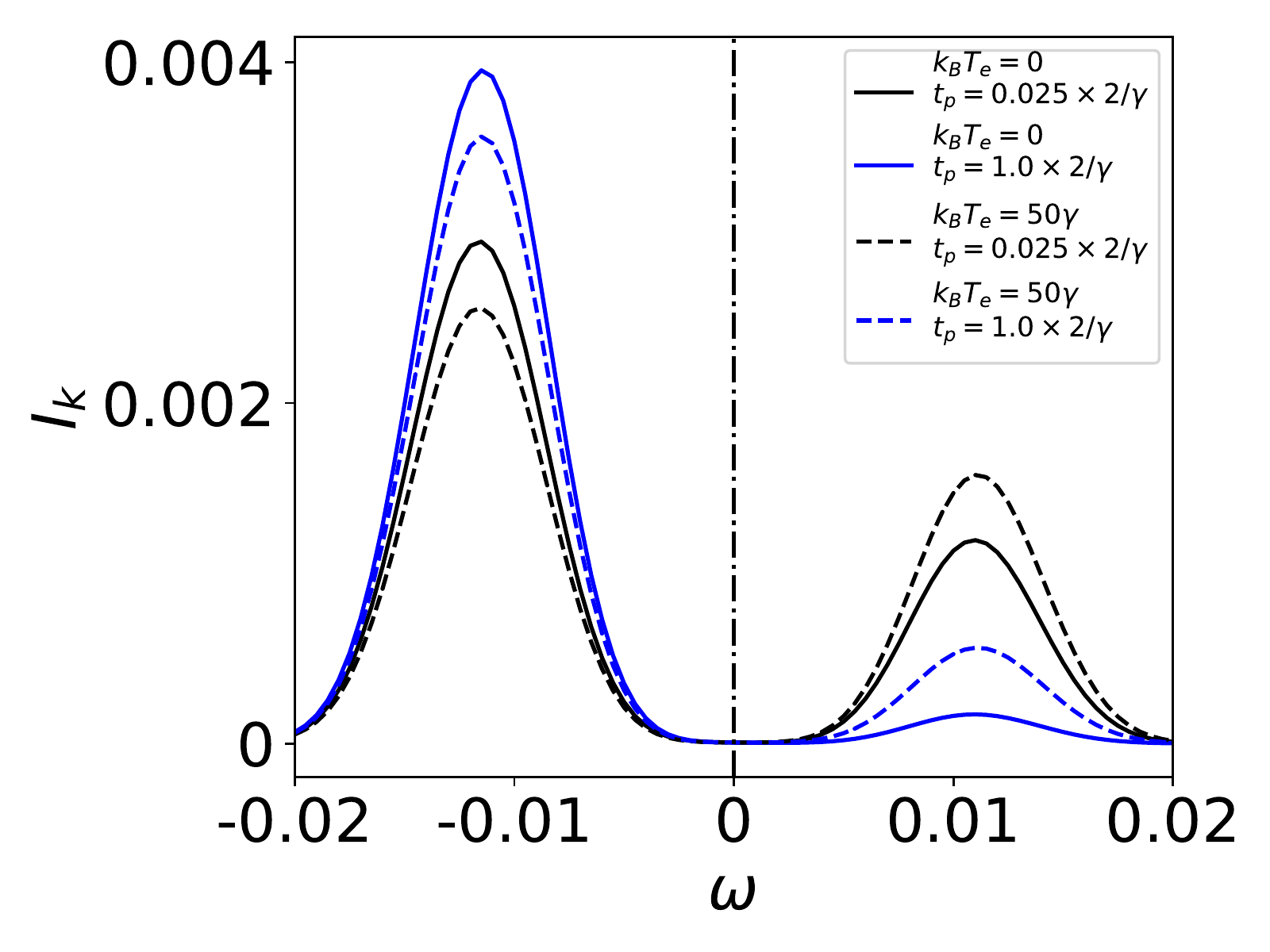}
    \caption{Non-equilibrium calculations with BCS order parameter featuring constant with a quench, as mentioned in \eq{qgap}, with $\Delta=0.01$.
    Upper panel: tr-ARPES signals with $\epsilon_k=k$, $\gamma=0.0001$, and $\sigma=400=0.02\times2/\gamma$. The two different $t_p$'s, short and long time after quench, are indicated in the figures. Lower panels: $I_\omega(k,t)$ (left) and $I_k(\omega,t)$ (right) at $k_BT_e=0$ (solid line) and $k_BT_e=50\gamma$ (dash line), both at short (black line) and long (blue line) time after quench. The vertical dash-dotted lines show where $k$ (left) or $\omega$ (right) is zero.}
    \label{fig:quenchT}
\end{figure}

With this new $\Sigma^<$, one can describe $G^<$ of electrons whose temperature is increased by the pump field. Such non-zero $T_e$ smears out $I(k,\omega,t)$ in both $k$ and $\omega$ directions. We demonstrate this effect by presenting $I(k,\omega,t)$ at a different temperature of three systems: a metallic system ($\Delta=0$), see \fig{metalT}; a BCS system ($\Delta=0.01$), see \fig{BCST}; and one of the non-equilibrium systems that we have considered in the main text (\eq{qgap}), see \fig{quenchT}.

By integrating $I(k,\omega,t)$ over $k$ or $\omega$, we calculate two quantities, $I_\omega(k,t)$ and $I_k(\omega,t)$:
\begin{equation}
    I_\omega(k,t)=\int d\omega I(k,\omega,t),
\end{equation}
and
\begin{equation}
    I_k(\omega,t)=\int dk I(k,\omega,t).
\end{equation}
These two quantities allow us to directly compare tr-ARPES signals among different temperatures.

Figures (\ref{fig:metalT}) and (\ref{fig:BCST}) clearly show that, as temperature increases, the lower energy states are less occupied, while the higher states are more occupied. This happens both in $k$ and $\omega$. On the other hand, \fig{quenchT} shows that both a non-equilibrium (i.e. time-varying) order parameter and a non-zero $T_e$ may change occupation of the energy states. However, the relaxation introduced by $\gamma$ eventually causes relaxation of the excitations from non-equilibrium order parameter.

\clearpage
\bibliography{cmp}

\end{document}